\documentclass[pdflatex,sn-mathphys-num]{sn-jnl}

\usepackage{graphicx}%
\usepackage{multirow}%
\usepackage{amsmath,amssymb,amsfonts}%
\usepackage{amsthm}%
\usepackage{mathrsfs}%
\usepackage[title]{appendix}%
\usepackage{xcolor}%
\usepackage{textcomp}%
\usepackage{manyfoot}%
\usepackage{booktabs}%
\usepackage{algorithm}%
\usepackage{algorithmicx}%
\usepackage{algpseudocode}%
\usepackage{listings}%
\usepackage{tikz}%
\usepackage{mathtools}%

\theoremstyle{thmstyleone}%
\theoremstyle{thmstyletwo}%

\theoremstyle{thmstylethree}%

\newcommand \bra[1] {\left\langle {#1} \right\vert}
\newcommand \ket[1] {\left\vert {#1} \right\rangle }
\newcommand \braket[2] {\left\langle {#1} \vert{#2} \right\rangle}

\begin{document}


\title[Article Title]{On the Symplectic Propagation of the Spin-MInt Algorithm for Non-Adiabatic Quantum Dynamics}


\author[1]{\fnm{James R.} \sur{Rampton}}

\author[1]{\fnm{Lauren E.} \sur{Cook}}

\author*[1]{\fnm{Timothy J. H.} \sur{Hele}}\email{t.hele@ucl.ac.uk}

\affil[1]{Department of Chemistry, University College London, Christopher Ingold Building, London WC1H 0AJ, United Kingdom}


\abstract{Mapping methods are often used for the numerical simulation of nonadiabatic systems by propagating classical mapping variable trajectories. A recently popularised mapping method is spin-mapping, whose mapping variables arise from quantum mechanical operators with symmetries described by a Lie-Poisson algebra. Simulating the classical-like dynamics of spin-mapping systems accurately is generally challenging, with many methods unable to preserve the underlying geometric structure of the symplectic form. The Spin-MInt algorithm is a recently proposed algorithm propagating spin-mapping variables, with a direct proof of symplecticity existing only for 2 electronic states. Here, we directly prove the symplecticity of the Spin-MInt algorithm for a general $K$ electronic states. A review of the symplectic nature of coadjoint orbits of the $\mathfrak{su}(K)$ Lie-Poisson algebra provides the framework needed to understand symplecticity of the Spin-MInt algorithm in this general case. The symplecticity of the method on the associated coadjoint orbit is then shown for what we believe to be the first time via an explicit verification of the symplecticity condition $\mathbf{MJ}\mathbf{M}^\textrm{T}=\mathbf{J}$ exploiting the Lie-Poisson structure of the system. To our knowledge, this is the first time the monodromy matrix for the Spin-MInt algorithm has been explicitly stated using canonical coordinates on the coherent state manifold for a general number of states. We hope that this will assist the development of classical-like spin-mapping methods which might utilise elements of the monodromy matrix, and inform future work on similar symplectic algorithms for coupled and uncoupled Lie-Poisson systems.}

\keywords{Symplecticity, Spin-Mapping, Nonadiabatic, Dynamics}



\maketitle

\section{Introduction}\label{secIntro}

One area of chemistry where the simulation of Hamiltonian dynamics is extremely important is in the field of nonadiabatic dynamics, which is useful for the study of many ultrafast energy and charge transfer processes \cite{zhu_charge_2015,cheng_dynamics_2009,marcus_electron_1985,runeson_exciton_2024,muh_refined_2012,domcke_role_2012,hammes-schiffer_theory_2010}. Approaches to the simulation of nonadiabatic systems are varied, ranging from extremely accurate but computationally expensive methods such as Multi-Configurational Time-Dependent Hartree-Fock (MCTDH) \cite{beck_multiconfiguration_2000,shin_multiple_1996,wang_multilayer_2003,mukherjee_assessing_2025,van_haeften_propagating_2023} to much cheaper yet more approximate methods such as Nonadiabatic extensions of Ring Polymer Molecular Dynamics (RPMD) \cite{ananth_mapping_2013,richardson_communication_2013,hel11a}, Non-adiabatic Matsubara Dynamics (NA-Mats) \cite{hele_boltzmann-conserving_2015,chowdhury_non-adiabatic_2020}, as well as Initial Value Representation (IVR) based approaches \cite{van_vleck_correspondence_1928,herman_semiclasical_1984,miller_semiclassical_2001,church_nonadiabatic_2018,kay_hermankluk_2006,sun_forwardbackward_1999}. There are also several mixed quantum-classical approaches with varying associated costs and accuracies, including the mean field based Ehrenfest methods \cite{atsango_accurate_2023,shalashilin_multiconfigurational_2011,tully_perspective_2012,zimmermann_efficient_2014,lieberherr_two-dimensional_2025}, surface hopping approaches \cite{kapral_surface_2016,mannouch_mapping_2023,shakib_ring_2017,tully_perspective_2012,tully_trajectory_1971,zimmermann_efficient_2014}, and mapping methods \cite{meyer_classical_1979,mannouch_mapping_2023,stock_classical_2005,bossion_non-adiabatic_2021,stock_semiclassical_1997,runeson_spin-mapping_2019,cotton_symmetrical_2013,cook_spin-mint_2026, cook_which_2023}.

Of these, in this article we focus on the mapping approach which maps quantum electronic systems onto associated classical phase space variables which are subsequently propagated according to Hamiltonian dynamics \cite{thoss_mapping_1999,stock_classical_2005}. Mapping methods can be utilized to investigate dynamical properties of nonadiabatic systems through the approximate calculation of time-correlation functions of two observable operators. In theory the exact quantum time-correlation functions would be calculated by integrating over all possible paths in the phase space, which is unfeasible for a numerical method \cite{feynman_quantum_2010,miller_classical_1970,gutzwiller_periodic_1971,hele_deriving_2016,althorpe_non-adiabatic_2016}. Consequently a great number of approximate time-correlation functions have been developed which aim to approximate, at varying degrees of computational cost and accuracy, the exact quantum time-correlation function. Some methods such as the Linearised Semi-Classical (LSC)-IVR \cite{ananth_semiclassical_2007,miller_semiclassical_2001,miller_electronically_2009,sun_semiclassical_1998,sun_semiclassical_19981,wang_semiclassical_1998} contain a phase factor which can make convergence challenging. Methods to circumvent this additional challenge often exploit a prefactor which is a function of elements of the monodromy matrix (the Jacobian of the time propagation map associated to the numerical method) \cite{moscato_time_2025,herman_semiclasical_1984,ananth_semiclassical_2007,miller_semiclassical_2001}. Consequently, highly accurate trajectories and accurate monodromy matrix computation is of great importance to converge these calculations, and therefore symplectic algorithms are desirable due to their high propagation accuracy \cite{craig_quantum_2004,ananth_mapping_2013,cook_electronic_2025,hele_boltzmann-conserving_2015}. 

Some of the more popular mapping methods are given by the Meyer-Miller-Stock-Thoss (MMST) and spin-mapping representations. The MMST representation describes the $K$ levels of a $K$-level electronic system as $K$ distinct 1-dimensional harmonic oscillators, each with an associated classical position $\mathbf{q}$ and momenta $\mathbf{p}$ \cite{meyer_classical_1979,stock_semiclassical_1997, ananth_mapping_2013, richardson_communication_2013, richardson_analysis_2017, church_nonadiabatic_2018,hele_deriving_2016}. This representation lends itself very favourably to treatment with the canonical classical phase space $\mathbf{q},\mathbf{p} \in \mathbb{R}^{2K}$ which led to the development of the MInt algorithm \cite{church_nonadiabatic_2018,cook_which_2023} for the symplectic propagation of the coupled electronic mapping variables and classical nuclear variables.

More recently, the spin-mapping representation has seen a surge in popularity and promising results surrounding the efficiency and accuracy of spin-mapping related methods have been observed \cite{richardson_nonadiabatic_2025,runeson_spin-mapping_2019, runeson_generalized_2020,runeson_quantum_2021,bossion_non-adiabatic_2021,bossion_non-adiabatic_2022,mannouch_partially_2020,mannouch_mapping_2023}. However, as far as we are aware, a detailed account of the formal mathematical background of spin-mapping has not yet been explicitly given, particularly how spin-mapping may be formally derived from considering only the associated symmetry group and thorough investigations of the associated phase spaces. The spin-mapping representation considers the $K$-level electronic system through its symmetry group, the Lie group $SU(K)$ and corresponding Lie algebra $\mathfrak{su}(K)$. The resulting phase-space is unlike the standard Euclidean phase space obtained in the MMST representation \cite{runeson_generalized_2020} and hence, investigating symplecticity within this phase space requires great care. 

In particular, the spin-mapping phase space for a closed, $K$-level electronic system is the complex projective space $\mathbb{CP}^{K-1}$, which introduces several layers of complexity that are not present when considering a typical $\mathbb{R}^{2n}$ phase space. The complex projective space $\mathbb{CP}^{K-1}$ may be thought of as the space $\mathbb{C}^{K}$ where two points $z_1,z_2 \in \mathbb{C}^K$ are identified if they lie on the same complex plane passing through the origin, or equivalently, if $z_1 = c z_2$ for some $c \in \mathbb{C}$. Due to the canonical coordinates on $\mathbb{CP}^{K-1}$ being slightly unintuitive \cite{bengtsson_geometry_2017,oh_action-angle_1994}, and there existing an embedding back into the Lie algebra $\mathfrak{su}(K)$ with desirable properties, dynamics are often treated with a set of over-complete global coordinates $\mathbf{u}$ defined by the Lie algebra $\mathfrak{su}(K)$. As a result, it is difficult to verify symplecticity at the level of the monodromy matrix since this matrix is inherently understood in local coordinates. 

The first symplectic propagator for the spin-mapping representation was recently proposed \cite{cook_spin-mint_2026}, where a direct proof of symplecticity was presented in the case for $K=2$. In this case, one has the isomorphism $\mathbb{CP}^1 \cong \mathbb{S}^2$, where $\mathbb{S}^2$ denotes the sphere in 3-dimensional space. Using intuition based on the sphere, as well as properties specific to $SU(2)$, the proof is relatively straightforward. The Spin-MInt algorithm was also shown to be symplectic for a general number of levels through showing the equivalence to the MInt algorithm which is known to be symplectic \cite{cook_spin-mint_2026, cook_which_2023, church_nonadiabatic_2018}. This approach, however, does not explicitly determine the Monodromy Matrix or help illuminate the underlying geometric structure being preserved since, for the MInt algorithm, symplecticity is on the space $\mathbb{R}^{2(K+F)}$ for $F$ nuclear degrees of freedom, which does not naturally relate to the spin-mapping variables. This work instead seeks to extend the direct, explicit proof to an arbitrary number of electronic levels whilst providing appropriate background for determining symplecticity on the spin-mapping manifold $\mathbb{CP}^{K-1}\times \mathbb{R}^{2F}$.

More generally, the symplecticity of numerical integrators is typically established via geometrical arguments. In particular, symmetric composition methods are symplectic under exact propagation of Hamiltonians or sub-Hamiltonians \cite{leimkuhler_simulating_2005}. While this provides a powerful and general framework, if one were looking to verify the symplecticity of a proposed algorithm, it would have to be identified with such a Hamiltonian flow. For some algorithms, this identification is not straightforward, and the connection between the abstract geometrical structure and its numerical realisation can be difficult to exploit.

In the context of classical systems whose Poisson bracket structure may be obtained from a Lie bracket, often called Lie-Poisson systems, there exists a natural symplectic form associated to the system, or more specifically, the coadjoint orbits of the system. This symplectic structure is well understood using the Kirillov-Kostant-Souriau (KKS) symplectic form \cite{kirillov_unitary_1962,dudley_quantization_1970,Souriau1970} naturally given to a coadjoint orbit (a symplectic manifold associated to the Lie group). The Runge-Kutta-Munthe-Kaas (RKMK) method is a popular choice for numerical approaches to coadjoint orbits, but is in general only structure preserving, a weaker condition than being symplectic \cite{munthe-kaas_lie-butcher_1995,cucker_iterated_1997,munthe-kaas_runge-kutta_1998,munthe-kaas_high_1999,iserles_lie-group_2000}. In cases where the RKMK approach yields a symplectic integrator, this integrator requires implicit solutions \cite{lasagni_canonical_1988,sanz-serna_runge-kutta_1988,suris_canonicity_1989}, while the Spin-MInt method is fully explicit \cite{cook_spin-mint_2026}. Furthermore, even in these symplectic cases, symplecticity is established at a structural level, without explicitly verifying the property for the discrete flow map itself. In particular, the associated monodromy matrix is not constructed, impeding the implementation of such methods into larger schemes which require elements of the monodromy matrix such as LSC-IVR \cite{moscato_time_2025,herman_semiclasical_1984,ananth_semiclassical_2007,miller_semiclassical_2001}.

Instead, the present work provides a direct, constructive proof of symplecticity at the level of the monodromy matrix for a class of Lie algebraic integrators. This approach makes explicit the use of Lie algebraic structures, canonical coordinates on non-Euclidean phase spaces and embeddings into adjoint spaces. We demonstrate this approach in the context of Spin-Mapping, providing justification for the symplecticity of the generalised multi-state version of the recently proposed Spin-MInt algorithm \cite{cook_spin-mint_2026}. Importantly, the proof is self-contained and does not rely on the equivalence to previously established symplectic numerical integrators, but verifies the property directly for the Spin-MInt integrator. More broadly, this work provides a systematic framework for verifying the symplecticity of numerical integrators on the coadjoint orbits of Lie-Poisson manifolds through the algebraic derivation of the monodromy matrix. Investigating the result of this explicit calculation can also provide insight into where a non-symplectic method is failing to be symplectic. By investigating the origins of discrepancies between the result of the explicit calculation and the symplectic structure matrix, one may be able to reverse engineer a symplectic integrator.

The structure of this article is as follows: In Sec. \ref{sec2} we provide the background theory of Lie-Poisson systems, how they apply to spin-mapping, the symplectic manifolds associated with them and also give the details of the Spin-MInt algorithm. In Sec. \ref{sec3} we give an overview of how the mathematical formalism relates specifically to spin-mapping and set up the proof of symplecticity. An outline of the full explicit direct proof is presented in Sec.~\ref{sec4}, further detailed in Appendix \ref{secA1}. We then discuss a possible generalisation prompted by the mathematical formalisation and some further uses of this proof method in Sec. \ref{secUses} and finally conclude in Sec. \ref{secConc}.

\section{Background Theory}\label{sec2}

Spin-mapping is used to model the classical-like dynamics generated by mapping a $K$-level quantum electronic system onto a classical phase space. In general, a $K$-level system consists of $K$ electronic states, each of which may be coupled to each other. Spin-mapping was originally derived from the equivalence between a 2-level system and a spin-1/2 particle, which are both described by $SU(2)$ \cite{sakurai_modern_2020,runeson_spin-mapping_2019}. However, when one considers a $K$-level system instead, the natural symmetry group is that of $SU(K)$ \cite{bertlmann_bloch_2008,runeson_generalized_2020,fano_symmetries_1996}, rather than the generalisation of a spin-$j$ system which is a $2j+1$ dimensional representation of $SU(2)$ \cite{fano_symmetries_1996,wightman_einige_1993}.

The electronic $K$-level system is also coupled to a nuclear system with $F$ degrees of freedom, however, this nuclear system is considered at the classical level with the standard phase space $(\mathbf{R},\mathbf{P})\in\mathbb{R}^{2F}$. We will assume that the Hamiltonian operator for this system takes the form \cite{runeson_spin-mapping_2019}
\begin{align}\label{Hamiltonian}
    \mathbf{H}(\mathbf{P},\mathbf{R}) = \frac{1}{2}\mathbf{P}^\textrm{T} \boldsymbol{\mu}^{-1}\mathbf{P} ~\mathbf{I} + \mathbf{V}(\mathbf{R})
\end{align}
for $\mathbf{V}(\mathbf{R})$ the $K\times K$ diabatic potential matrix, $\mathbf{I}$ the $K\times K$ identity matrix, and $\boldsymbol{\mu}$ the $F\times F$ diagonal matrix of nuclear masses. For some representations, careful separation of an identity component of $\mathbf{V}(\mathbf{R})$ is desired to improve computational properties, which is normally written as an additional term $U(\mathbf{R})\mathbf{I}$  \cite{muller_flow_1999,cotton_trajectory-adjusted_2019}, and can sometimes yield different results \cite{saller_identity_2019, richardson_communication_2013, cook_which_2023}. The spin-mapping representation fortunately does not suffer from this problem as the natural separation into traceless and identity-like terms provided by the $SU(K)$ Lie group described later does this automatically  \cite{runeson_generalized_2020}, hence we do not separate a $U(\mathbf{R})\mathbf{I}$ term explicitly.

For a $K$-level electronic system, all electronic observable operators are represented by $K \times K$ Hermitian matrices which can be complex. Such operators act on electronic states defined by arbitrary normalised superpositions of the basis states $\ket{n}$, where $n \in \{1,\dots,K\}$. The Hamiltonian operator is an example of this and is often separated into two components, one proportional to the identity operator generating trivial electronic evolution, and a traceless Hermitian component. The traceless Hermitian component lies in the Lie algebra $i~\mathfrak{su}(K)$, where $\mathfrak{su}(K)$ denotes the $K^2-1$ dimensional Lie algebra of traceless skew-Hermitian matrices and $i = \sqrt{-1}$ the imaginary unit. The Lie algebra $\mathfrak{su}(K)$ governs the non-trivial unitary dynamics of observable matrix valued operators. Since $\mathfrak{su}(K)$ is a real Lie algbera \cite{fulton_representation_2004}, the behaviour of $\mathfrak{su}(K)$ and $i~\mathfrak{su}(K)$ are identical up to the imaginary constant, so we can consider each interchangeably for the most part. Therefore, we will only include the factor $i$ when it is important to distinguish the two cases. In studying dynamics, one seeks to characterise the time evolution of relevant observables. In quantum mechanics, this is achieved via exponentiation of the Hamiltonian to obtain a time evolution operator, resulting in an element of the associated Lie group $SU(K)$\cite{hall_lie_2015}.

To emphasise the difference between the Lie group $SU(K)$ and Lie algebra $\mathfrak{su}(K)$, we give the definition of each. The Lie group is defined as
\begin{align}
    SU(K) = \{ \mathbf{Y}\in M_{K\times K}(\mathbb{C}) : \mathbf{Y}\mathbf{Y}^\dagger = \mathbf{I} \text{ and } \det(\mathbf{Y}) = 1 \},
\end{align}
while the Lie algebra is defined as
\begin{align}
    \mathfrak{su}(K) = \{ \mathbf{X}\in M_{K\times K}(\mathbb{C}): -\mathbf{X} = \mathbf{X}^\dagger \text{ and } \text{Tr}[\mathbf{X}] = 0\}.
\end{align}
From these properties, one can quickly see that exponentiating elements of the Lie algebra results in elements of the Lie group since $\mathbf{Y}^\dagger = (e^\mathbf{X})^\dagger = e^{(\mathbf{X}^\dagger)} = e^{-\mathbf{X}} = \mathbf{Y}^{-1}$ and $\det(\mathbf{x}) =  \det{(e^\mathbf{Y})} = e^{\text{Tr}[\mathbf{Y}]} = 1$.

\subsection{Lie-Poisson Algebras}

The problem of studying $K$-level electronic systems is therefore naturally presented as that of studying the Lie algebra $\mathfrak{g} = \mathfrak{su}(K)$ and how it acts on the Hilbert space $\mathcal{H} = \mathbb{C}^K$. Formally a Lie algebra is a vector space $\mathfrak{g}$ together with an alternating bilinear map $[\cdot,\cdot]:\mathfrak{g}\times\mathfrak{g}\rightarrow\mathfrak{g}$ which satisfies the Jacobi identity \cite{hall_lie_2015}, however, since $\mathfrak{su}(K)$ is associative, the commutator $[x,y] = xy-yx$ satisfies all requirements for the Lie bracket \cite{humphreys_introduction_1972}. We will therefore assume that the commutator plays the role of the Lie bracket for all Lie algebras within this study. Since $\mathfrak{g}$ is a vector space, it has a basis $\lambda_i$ which, through the Lie bracket, defines the 3rd rank tensor of structure constants $f_i{}_j{}^k$ via $[\lambda_i,\lambda_j] = if_i{}_j{}^k \lambda_k$ \cite{bossion_general_2021}. Throughout this article Einstein summation convention is used implicitly summing over repeated raised and lowered indices and we define the structure constants to be real by introducing the factor $i$.

To consider classical-like dynamics, we first require a sensible mapping from the space of operators (the Lie algebra $\mathfrak{g}$) to the field $\mathbb{R}$. This is formalised through the notion of a dual space, $\mathfrak{g}^*$. The dual space $\mathfrak{g}^*$ is defined to be the set of linear functionals on $\mathfrak{g}$,
\begin{align}
    \mathfrak{g}^* = \{u : \mathfrak{g}\rightarrow\mathbb{R} : u \text{ is linear}\}.
\end{align}
This space is a vector space of dimension equal to $\mathfrak{g}$ \cite{katznelson_linalg_2008}, and hence has a basis $\xi^k \in \mathfrak{g}^*$. We will use the raised index for the basis of $\mathfrak{g}^*$ so that elements of $\mathfrak{g}$ may be easily distinguished from those of $\mathfrak{g}^*$. Further motivation of using raised and lowered indices may be found in ref. \cite{spivak_comprehensive_1979}. This dual space therefore constitutes the setting in which classical-like mechanics are most naturally formulated \cite{marsden_introduction_1999}.

The way one defines the dynamics on a classical system is through the Poisson bracket \cite{poisson_memoire_1809}, which for a normal position and momentum based phase space $\mathbb{R}^{2n}$ takes in two smooth functions, $g,h: \mathbb{R}^{2n} \rightarrow \mathbb{R}$ and returns a third smooth function, $\{g,h\} : \mathbb{R}^{2n}\rightarrow\mathbb{R}$ defined by
\begin{align}
    \{g,h\} = \sum_{i=1}^n \left( \frac{\partial g}{\partial q_i}\frac{\partial h}{\partial p_i} - \frac{\partial g}{\partial p_i}\frac{\partial h}{\partial q_i} \right).
\end{align}
Here we use $n$ for the dimension of an arbitrary phase space which should not be confused with the variables $F$ and $K$ reserved for the spaces associated with spin-mapping. In general, the dual of a Lie algebra does not have such position and momentum coordinates, meaning that this definition of the Poisson bracket does not apply.

Instead, one looks to construct the Poisson bracket by exploiting the structure provided by the Lie bracket. To mimic the notion of differentiation present in the normal Poisson bracket on the dual space, one needs the exterior derivative \cite{marsden_semidirect_1984}. For a smooth function $g:\mathfrak{g}^*\rightarrow\mathbb{R}$ (henceforth denoted $g \in C^\infty(\mathfrak{g}^*)$), one defines the exterior derivative as the linear map $dg:T_u\mathfrak{g}^*\rightarrow \mathbb{R}$ that best approximates $g$ at $u \in \mathfrak{g}^*$ \cite{spivak_comprehensive_1979}. The space $T_u\mathfrak{g}^*$ is the tangent space of $\mathfrak{g}^*$ at $u$, consisting of all the vectors tangential to $\mathfrak{g}^*$ at $u$. In general, the tangent space is important as the base space may not have a notion of addition and thus a linear function is not well defined. However, since $\mathfrak{g}^*$ is already a vector space, the tangent space may be identified with the base space $T_u\mathfrak{g}^*\cong\mathfrak{g}^*$ \cite{lee_introduction_2003}. Now since $dg:T_u\mathfrak{g}^* \rightarrow\mathbb{R}$ is a linear functional, we may interpret it as an element of a further dual space, $dg \in (T_u\mathfrak{g}^*)^* \cong (\mathfrak{g}^*)^* \cong \mathfrak{g}$ since a dual of a dual may be identified with the original space \cite{halmos_finite-dimensional_1974}. The object $dg$ is often referred to as a 1-form since it takes 1 element of $T_u\mathfrak{g}^*$ and returns a real scalar. Note that due to $\mathfrak{g}^*$ being the central space of study, we say 1-forms $dg\in \mathfrak{g}$ rather than the typical definition of a 1-form $dg \in \mathfrak{g}^*$ when $\mathfrak{g}$ is the central space of study.

Now that we have a differentiated object which may be identified with an element of the Lie algebra and so has a well-defined antisymmetric product, we can define the Poisson bracket of the Lie algebra $\{\cdot,\cdot\}:C^{\infty}(\mathfrak{g}^*)\times C^{\infty}(\mathfrak{g}^*) \rightarrow C^\infty(\mathfrak{g}^*)$ by how it acts on an arbitrary $u\in\mathfrak{g}^*$ \cite{marsden_introduction_1999}:
\begin{align}
    \{ g,h \}(u) := u([dg,dh]).
\end{align}
This expression can also be written in coordinates. A point $u \in \mathfrak{g}^*$ may be written $u = u_k\xi^k$, which allows us to write
\begin{align}
    dg = \frac{\partial g}{\partial u_k} du_k
\end{align}
where $du_k$ is the derivative of the map giving the $k$th coordinate, which is precisely the $k$th element of the basis of $(T_u\mathfrak{g}^*)^*$ \cite{abraham_manifolds_1988}. Since $(T_u\mathfrak{g}^*)^* \cong \mathfrak{g}$, we may identify $du_k \leftrightarrow \lambda_k$ giving 
\begin{align}
    \{g,h\}(u) := u([dg,dh]) \leftrightarrow u_k \lambda^k\left(\frac{\partial g}{\partial u_i}\frac{\partial h}{\partial u_j} [\lambda_i,\lambda_j]\right) = i{f_{ij}}^k u_k \frac{\partial g}{\partial u_i}\frac{\partial h}{\partial u_j},
\end{align}
where $\lambda^k$ denotes the basis dual to $\lambda_l$, such that $\lambda^k(\lambda_l) = \delta^k{}_l$. For the remainder of this work, we will not write identifications between $T_u^*\mathfrak{g}^*$ and $\mathfrak{g}$ explicitly and just use the equality symbol instead. We will also assume that $\mathfrak{g}^*$ always uses the dual basis to that of $\mathfrak{g}$.

Considering a Hamiltonian function $H\in C^\infty(\mathfrak{g}^*)$, and letting $g = iH$, $h=u_j$, we can find the dynamics of the coordinates of $u$ generated by $H$,
\begin{align}\label{LPDynamics}
    \dot{u}_j :=i\{H,u_j\}  = -f_i{}_j{}^k u_k \frac{\partial H}{\partial u_i},
\end{align}
or since $f_i{}_j{}^k$ is totally antisymmetric, we have the more commonly used version,
\begin{align}
    \dot{u}_i = f_i{}_j{}^ku_k \frac{\partial H}{\partial u_j}.
\end{align}
In the case of the three dimensional $\mathfrak{g} = \mathfrak{su}(2)$ with $f_{ij}{}^k \propto \epsilon_{ij}{}^k$ the levi-civita symbol, this can be rewritten as the cross product $\dot{\mathbf{u}} = \mathbf{H}\times \mathbf{u}$, which is precisely Heisenberg's equation of motion for the spin-vector $\mathbf{u}$ \cite{bloch_nuclear_1946,runeson_spin-mapping_2019,bossion_non-adiabatic_2021, cook_spin-mint_2026}. In taking $g=iH$, we have assumed the quantum propagator is defined using $\hbar=1$, which we will continue to do throughout this work.

However, thus far we only have dynamics on the dual space $\mathfrak{g}^*$. Considering specifically the Lie algebra $i~\mathfrak{su}(K)$, it may naturally be given an inner product
\begin{align}\label{TrInnerProd}
    \langle \mathbf{X},\mathbf{Y}\rangle_{\mathrm{Tr}} = \frac{1}{2}\text{Tr}[\mathbf{X}\mathbf{Y}].
\end{align}
This provides an identification between the dual space $\mathfrak{su}(K)^*$ and the original space $\mathfrak{su}(K)$ through the non-degenerate \cite{hall_lie_2015} map
\begin{align}
    \phi_{\mathbf{X}}(\mathbf{Y}) = \langle\mathbf{X},\mathbf{Y}\rangle_\mathrm{Tr},
\end{align}
which fully describes the action of every element $\phi$ of the dual space $\mathfrak{su}(K)^*$. Therefore these Lie-Poisson dynamics formulated on $\mathfrak{g}^*$ are not the desired classical-like dynamics since they still describe the evolution of operators rather than classical variables.

\subsection{Spin-Mapping}

A mapping representation is one where quantum operators are mapped onto classical variables which can then be propagated while maintaining quantum properties. Spin-mapping in particular defines the classical variable $\mathbf{u}\in\mathbb{R}^{K^2-1}$ arising from a $K$-level system by recreating the Lie-Poisson dynamics associated to the $K^2-1$ dimensional Lie algebra $\mathfrak{su}(K)$. The classical-like dynamics of the variable $\mathbf{u}$ are understood through the lens of the coherent state. Since spin-mapping focuses on a $K$-level system, the relevant coherent state is given by an arbitrary normalised superposition of each of the individual states $\ket{1},\dots,\ket{K}$ \cite{perelomov_generalized_1986}. We will write this superposition
\begin{align}\label{CoherentState}
    \ket{\Omega} = \sum_{n=1}^K \ket{n}\braket{n}{\Omega} := \sum_{n=1}^K \cos\frac{\theta^n}{2}\prod_{l=1}^{n-1}\sin\frac{\theta^l}{2}e^{i\phi^l} \ket{n},
\end{align}
for generalised Euler angles  $\theta^i \in [0,\pi], \phi^i \in [0,2\pi]$ for indices $i \in \{1,\dots,K-1\}$ and where $\theta^K := 0$ and a sum or product from $l$ to $l-1$ is equal to zero or one respectively. Such a state $\ket{\Omega}$ is often called the spin coherent state \cite{runeson_generalized_2020,bossion_non-adiabatic_2022}.

The angular variables parameterising the spin coherent state are chosen in concordance with previous literature \cite{runeson_generalized_2020,bossion_non-adiabatic_2022} as a set of coordinates on the associated phase space manifold $\mathbb{CP}^{K-1}$. There are other possible choices of coordinates for this manifold, including the commonly used stereographic coordinates facilitating easier exploitation of the K{\"a}hler (complex) structure \cite{bengtsson_geometry_2017,oh_action-angle_1994,barnes_cpn_2002} and some slightly differently defined angular variables \cite{bengtsson_geometry_2017} resulting in the action angle variables defined in \cite{oh_action-angle_1994}. Justification of why the manifold $\mathbb{CP}^{K-1}$ was selected will be presented in Sec.~\ref{subsecChooseLeaf}.

To connect the Lie-Poisson and classical-like dynamics, a mapping from operator space to phase space is required. For Euclidean phase spaces such as that associated with the MMST representation, this role is typically filled by the Wigner transform, mapping trace class operators $A$ to integrable functions $[A]_w:\mathbb{R}^{2n}\rightarrow\mathbb{R}$. However, this construction relies on vanishing curvature of the phase space and is therefore not directly applicable to $\mathbb{CP}^{K-1}$ \cite{brendle_classification_2008}.

Instead, we use the vector space structure of the Lie algebra $\mathfrak{su}(K)$. In particular we define the Generalised Gell-Mann (GGM) matrices $\lambda_i$ \cite{gell-mann_symmetries_1962,hioe_n_1981} which give a basis of $\mathfrak{su}(K)$. The GGM matrices are defined by
\begin{subequations}\label{GGMBasis}
\begin{align}
    \boldsymbol{\lambda}_{\alpha_{nm}} &= \mathbf{E}_{nm} + \mathbf{E}_{mn}, \\
    \boldsymbol{\lambda}_{\beta_{nm}} &= i(\mathbf{E}_{mn} - \mathbf{E}_{nm}), \\
    \boldsymbol{\lambda}_{\gamma_n} &= \sqrt{\frac{2}{n(n-1)}} \left(  (1-n)\mathbf{E}_{nn} + \sum_{l=1}^{n-1} \mathbf{E}_{ll}  \right),
\end{align}
\end{subequations}
where $\mathbf{E}_{nm}$ is the matrix with a 1 in position $(n,m)$ and 0 everywhere else and $\alpha,\beta,\gamma$ give the indices of the symmetric, anti-symmetric and diagonal matrices respectively as follows (where $n>m$):
\begin{subequations}
\begin{align}
    \alpha_{nm} &= n^2 + 2(m-n)-1,\\
    \beta_{nm} &= n^2 + 2(m-n),\\
    \gamma_n &= n^2-1.
\end{align}
\end{subequations}
This basis is orthonormal with respect to the trace inner product defined in Eqn.~\eqref{TrInnerProd}.

Using this basis, the Stratonovich-Weyl (SW) transform may be defined, providing a mapping of linear operators $\mathbf{B} \in \mathfrak{su}(K)$ onto functions $[\mathbf{B}]_s:\mathbb{CP}^{K-1}\rightarrow\mathbb{R}$. The SW transform is defined as
\begin{align}\label{SWTransform}
    [\mathbf{B}]_s(\mathbf{\Omega}) = \mathcal{B}_0+2r_su_i(\mathbf{\Omega})\mathcal{B}^i,
\end{align}
where $\mathbf{\Omega}$ denotes the vector $(\theta_1,\dots,\theta_{K-1},\phi_1,\dots,\phi_{K-1})$, $u_i: \mathbb{CP}^{K-1}\rightarrow \mathbb{R}$ is defined by the moment map
\begin{align}\label{PureMoMap}
u_i(\mathbf{\Omega}) = \bra{\Omega}\boldsymbol{\lambda}_i\ket{\Omega},
\end{align}
and the SW radius $r_s$ is a constant fixed depending on convention. $\mathcal{B}^i$ denotes components of $\mathbf{B}$ with respect to the trace inner product which can be written as
\begin{subequations}
\begin{align}
    \mathcal{B}^0 &= \frac{1}{K}\text{Tr}[\mathbf{B}\mathbf{I}], \\
    \mathcal{B}^i &= \frac{1}{2}\text{Tr}[\mathbf{B}\boldsymbol{\lambda}^i].
\end{align}
\end{subequations}
With these definitions, the decomposition $\mathbf{B} = \mathcal{B}^0 \mathbf{I} + \mathcal{B}^i\lambda_i$ is obtained, which is typical in spin-mapping approaches \cite{bossion_non-adiabatic_2021,bossion_non-adiabatic_2022,bossion_non-adiabatic_2023,runeson_generalized_2020,runeson_spin-mapping_2019}. The scaling constant $r_s$ and identity-like term $\mathcal{B}_0$ are introduced to allow for an improved degree of generality and to keep track of growth terms respectively, but neither play an important role in the dynamics. This mapping may be effectively understood as $\boldsymbol{\lambda}_i \mapsto 2r_su_i(\boldsymbol{\Omega}),~ \mathbf{I} \mapsto 1$.

\begin{figure}[h] \centering
\begin{tikzpicture}

    \draw (3,0) circle (2);
    \draw[dashed] (5,0) arc (0:180:2 and 0.6);
    \draw (1,0) arc (180:360:2 and 0.6);
    \draw[->] (3,0) -- (6,0)node[right]{$y$};
    \draw[->] (3,0) -- (3,3)node[above]{$z$};
    \draw[->] (3,0) coordinate (B) -- (1,-2)node[below]{$x$} coordinate (A);
    \draw[->][blue] (3,0) -- (3.6,1.6);
    \draw[dashed][blue] (3.6,1.6) -- (3.6,-0.3) coordinate (C);
    \draw[dashed][blue] (3.6,-0.3) -- (3,0);
    \draw[rotate around={7:(2.77,-0.23)}][blue] (2.77,-0.23) arc(220:330:0.3 and 0.15);
    \draw[rotate around={13.5:(1.42,1.2)}][red] (1.42,1.2) arc(210:330:1.45 and 0.45);
    \draw[dashed][rotate around={13.5:(3.85,1.8)}][red] (3.85,1.8) arc(30:150:1.45 and 0.45);
    \draw[->][red] (3,0) -- (2.6,2.58);
    \node at (2.6,2.78){\textcolor{red}{$\mathbf{H}$}};
    \node at (3,-0.41){\textcolor{blue}{$\phi$}};
    \draw[rotate around={-15:(3,0.9)}][blue] (3,0.9) arc(140:0:0.18 and 0.1);
    \node at (3.8,1.4){\textcolor{blue}{$\mathbf{u}$}};
    \node at (3.2,1.1){\textcolor{blue}{$\theta$}};
    \draw[<-] (3,0) -- (2.5,0.13);
    \draw[->] (2,0.26) -- (1.5,0.39);
    \node at (2.25,0.195){$r_s$};
    \draw[->] (3.6,1.6) -- (4.2,1.9);
    \node at (4.4,1.95){$\dot{\mathbf{u}}$};
    \node at (3.4,2.35){$\ket{1}$};
    \node at (3,-2.4){$\ket{2}$};

\end{tikzpicture}
\caption{A sketch of the phase space of the $SU(2)$ Lie group, with Euler angles and a Stratonovich-Weyl radius drawn on. We also include an arbitrary Hamiltonian with associated vector $\mathbf{H}$ in order to give an example of propagation of the $\mathbf{u}$ variable with $\dot{\mathbf{u}}$ perpendicular to both $\mathbf{H}$ and $\mathbf{u}$. This results in the circular precession of $\mathbf{u}$ depicted in red} \label{sphere}
\end{figure}
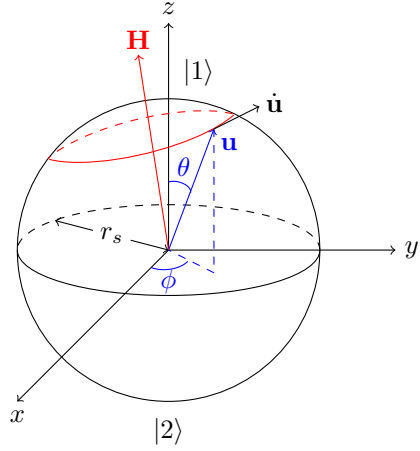

In the two-level case, this construction reduces to the standard embedding of spherical coordinates $(\theta,\phi)$ into $\mathbb{R}^3$ shown in Fig.~\ref{sphere}, reflecting the isomorphism $SU(2)\cong SO(3)$. For general $K$, such an isomorphism does not exist, so the geometric interpretation of this embedding is less transparent.

Through the extra structure provided by the GGM basis and the trace inner product \eqref{TrInnerProd}, the dual basis $\lambda^k$ can be understood as the map $\lambda^k:\mathfrak{g}\rightarrow\mathbb{R}$ defined by
\begin{align}
    \lambda^k(\mathbf{X}) = \text{Tr}[\boldsymbol{\lambda}_k \mathbf{X}]~~\forall \mathbf{X}\in \mathfrak{g}.
\end{align}
Due to $\text{Tr}[\boldsymbol{\lambda}_i\boldsymbol{\lambda}_j]$ being a well-defined metric relating the dual space to the original space, we can understand the operator which raises/lowers indices (the metric tensor) \cite{spivak_comprehensive_1979}, often denoted by $g_{ij}$, is precisely the identity matrix $\delta_{ij}$ in the GGM bases of $\mathfrak{su}(K)$ and $\mathfrak{su}(K)^*$. This means that in the case of the Lie-Poisson variables $u_i$ defined using the GGM matrices and other directly related variables, one can raise and lower indices freely, converting between geometric 1-forms and vectors without any additional calculation.

We note that the classical-like dynamics of the spin-vector $\mathbf{u}$ resulting from the SW transform are inherited entirely from the Lie-Poisson dynamics since the dynamics arise entirely from the choice of basis, which is consistent across both perspectives. However, the spin-vector variable is now a classical variable rather than an operator, so we have obtained the classical-like dynamics that we desired.



\subsection{Symplectic Leaves}\label{subsecSympLeaves}

Eqn.~\eqref{LPDynamics} describes entirely the dynamics of the electronic system, however, there is some additional structure masked beneath the equation in the form of orbits \cite{Kirillov_Orbit_2004}. One can show that depending on the value of $\mathbf{u} = (u_1,\dots,u_{N^2-1})$, there are some directions in which motion does not occur, and hence during propagation, the variable $\mathbf{u}$ may be constrained to a specific submanifold. These submanifolds are often referred to as the symplectic leaves \cite{weinstein_local_1983}.

To investigate the structure of these submanifolds further, we must formalise the notion of there being no motion in a given direction. To consider this motion, we first give an alternate definition of the Poisson bracket using the bilinear functional, sometimes called the Poisson bivector, $\pi_u : \mathfrak{g}\times\mathfrak{g} \rightarrow\mathbb{R}$ defined by $\pi_u(X,Y) = u([X,Y])$. Intuitively this should be thought of as the inverse of the symplectic form. One can define the motion associated to $X \in \mathfrak{g}$ by considering the unique \cite{kirillov_unitary_1962} element $v_X \in \mathfrak{g}^*$ such that for each $X \in \mathfrak{g}$,
\begin{align}
    v_X(Y) = \pi_u (X,Y) ~~ \forall Y \in \mathfrak{g}.
\end{align}
This gives the directions of motion at a point $u$ as precisely the space
\begin{align}
    T_u\mathcal{O}_u:= \{ v_X:X\in\mathfrak{g} \}.
\end{align}

To better understand this submanifold, a helpful map to study is the adjoint action $\text{ad}:\mathfrak{g}\times\mathfrak{g} \rightarrow\mathfrak{g}$ defined by $\text{ad}_XY = [X,Y]$. This in turn allows the definition of the coadjoint action, $\text{ad}^*:\mathfrak{g}\times \mathfrak{g}^*\rightarrow\mathfrak{g}^*$ by how it acts on an arbitrary $Y \in \mathfrak{g}, (\text{ad}_X^*u)(Y) = u([X,Y])$. Using this notation, one notices that $v_X = \text{ad}_X^*u$, so finding the directions in which there is no motion is equivalent to finding the stabiliser (sometimes referred to as the kernel) of the coadjoint action. One can define the stabiliser of the coadjoint action at $u \in \mathfrak{g}$ as 
\begin{align}
\text{Stab}_\mathfrak{g}(u) = \{ X\in \mathfrak{g} : u([X,Y]) = 0 ~~ \forall Y \in \mathfrak{g} \},
\end{align}
giving the natural isomorphism $T_u\mathcal{O}_u \cong \mathfrak{g}/\text{Stab}_\mathfrak{g}(u)$ where the $/$ represents taking the quotient group \cite{dummit_abstract_2004}. Loosely this quotient group is a copy of the Lie algebra where dimensions in which there is no motion have been removed. 

This construction also motivates the choice of notation $T_u\mathcal{O}_u$. Since these are directions of motion, they are inherently related to derivatives. To find the object they are a derivative of, we define the adjoint action instead at the Lie group level which is distinguished by the change in capitalisation
\begin{align}\label{OrbDef}
    \mathcal{O}_u = \{ \text{Ad}_g^*u : g\in G \}
\end{align}
where $\text{Ad}_g(X) = gXg^{-1}$ for $g\in G, X\in\mathfrak{g}$ and $(\text{Ad}_g^*u)(X) = u(\text{Ad}_{g^{-1}}X)$ for every $X \in \mathfrak{g}$. Then $\mathcal{O}\cong G/\text{Stab}_G(u)$ where
\begin{align}
    \text{Stab}_G(u) = \{ g\in G : (\text{Ad}_g^*u)(X) = u(X) ~~\forall X \in \mathfrak{g} \}.
\end{align}
where $G/\text{Stab}_G(u)$ is precisely the coadjoint orbit of $G$ at $u$.

At this point we remark that since the motion is never the 0 function on the manifold $T_u\mathcal{O}_u$, this is also precisely the manifold on which the map $X\mapsto v_X$ is invertible, giving a bijective correspondence between dual space and original space elements. This enables us to define the bilinear functional dual to the Poisson bivector, $\eta_u:\mathfrak{g}^*\times\mathfrak{g}^*\rightarrow \mathbb{R}$. This is sometimes referred to as a 2-form since it takes in 2 elements of $T_u\mathfrak{g}^*$, and it is non-degenerate on the submanifold $T_u\mathcal{O}_u$ \cite{kirillov_unitary_1962}. Since $ \mathfrak{g^*}\cong T_u\mathfrak{g}^*$, $\eta_u$ is a closed, non-degenerate, antisymmetric differential 2-form on $\mathfrak{g}^*$, which is precisely the definition of a symplectic form \cite{mcduff_introduction_2017}. A symplectic form constructed in such a way is often called a Kirillov-Kostant-Souriau (KKS) symplectic form \cite{kirillov_unitary_1962,dudley_quantization_1970,Souriau1970}.

Having the KKS symplectic form indicates that one can define classical Hamiltonian dynamics on each of the submanifolds $\mathcal{O}_u$, but also provides an idea for the testing of numerical methods. If the numerical propagation of $\mathbf{u}$ is constrained to the coadjoint orbit, then one can define the associated flow map $\phi$ as symplectic if it preserves the symplectic form $\eta_u(\phi(u),\phi(v)) = \eta_u(u,v)$.

\subsection{Canonical Coordinates}

The choice of coordinates one makes to describe the symplectic manifold is quite important. By Darboux's theorem for symplectic forms \cite{Darboux1882,Moser1965,mcduff_introduction_2017}, at every point $z$ in a symplectic manifold it is always possible to establish a local set of coordinates $\mathbf{q},\mathbf{p} \in \mathbb{R}^n$ where $n$ is the dimension of the arbitrary symplectic manifold, not to be confused with the spin-mapping specific variables $K$ or $F$. Note that these coordinates are currently general and not associated to any specific Hamiltonian. Using these coordinates, at the point $z$ one can write the symplectic form
\begin{align}
    \eta = \sum_{i=1}^n dq^i\wedge dp^i,
\end{align}
where $\wedge$ denotes the wedge product of two differential 1-forms \cite{lee_introduction_2003}. If we let the point $z$ be described by the local coordinates $\mathbf{z} = (\mathbf{q},\mathbf{p})$, this can equivalently be written
\begin{align}
    \eta = J_{ij} dz^i \wedge dz^j
\end{align}
where $\mathbf{J}$ is often called the symplectic structure matrix, which in the coordinates given by Darboux's theorem takes the form
\begin{align}
    \mathbf{J} = \begin{bmatrix}
        \mathbf{0} & \mathbf{I}\\
        -\mathbf{I} & \mathbf{0} 
    \end{bmatrix}.
\end{align}
where $\mathbf{0},\mathbf{I}$ represent the $n\times n$ zero and identity matrices respectively. We will call a set of coordinates that give this form of the structure matrix canonical coordinates, with $z_i, i \in \{1,\dots,n\}$ being canonical position coordinates and $z_i, i \in \{n+1,\dots,2n\}$ canonical momentum coordinates. We also consider the matrix representation $\boldsymbol{\Phi}$ of the linear map $d\phi$ for $\phi$ the flow map, sometimes called the monodromy matrix. Using this notation, we can rewrite the symplecticity condition as $\boldsymbol{\Phi} \mathbf{J} \boldsymbol{\Phi}^\textrm{T} = \mathbf{J}$.

However, sometimes the construction of these coordinates is non trivial. In this case, one can work in non canonical coordinates (although they should still have the correct dimensionality) by altering $\mathbf{J}$. If we instead consider the point $z$ in a $2n$ dimensional manifold with a symplectic form written in non-canonical coordinates,
\begin{align}
    \eta = \eta_{ij}(z) ~dz^i \wedge dz^j,
\end{align}
then for each point $z$, a $(1,1)$ tensor field $\Phi(z)$ with components $\Phi^i{}_j(z)$ is symplectic at $z$ if $\eta_{ij}(z) {\Phi^i}_k(z){\Phi^j}_l(z) = \eta_{kl}(z)$. Aiming for a form similar to the matrix condition, we define $J^{ij}(z)$ such that $J^{ik}(z)\eta_{kj}(z) = {\delta^i}_j$, then by taking the inverse of both sides, the condition becomes ${\Phi^k}_i(z) {\Phi^l}_j(z) J^{ij}(z) = J^{kl}(z)$, which to write more intuitively as a matrix multiplication, is ${\Phi^k}_i(z) J^{ij}(z) {\Phi_j}^l(z) = J^{kl}(z)$. Now removing the indices, the condition is $\Phi(z)J(z)\Phi^\textrm{T}(z) = J(z)$.

Therefore, in terms of the structure tensor $J$, one may equivalently show symplecticity by showing that $\Phi(z)J(z)\Phi^\textrm{T}(z) = J(z)$, or that $\Phi^\textrm{T}(z) J^{-1}(z)\Phi(z) = J^{-1}(z)$. By taking transposes of these equations, and noting that the symplectic form is necessarily antisymmetric so both $J^\textrm{T}=-J$ and $J^{-T}=-J^{-1}$, one may also show symplecticity by showing that
$\Phi^\textrm{T}(z) J(z)\Phi(z) = J(z)$ or $\Phi(z)J^{-1}(z)\Phi^\textrm{T}(z) = J^{-1}(z)$. In our proof, we opt to use canonical coordinates since they can be derived without much difficulty, but in general one may approach the problem looking to show that the map $\phi$ is symplectic for each $z$ in non-canonical coordinates.

In order to exploit various facets of the dynamics, we find that the most convenient form of the symplecticity criterion to show is
\begin{align}\label{sympCriterion}
    \boldsymbol{\Phi} \mathbf{J}^{-1}\boldsymbol{\Phi}^\mathrm{T} = \mathbf{J}^{-1}
\end{align}
and hence this is what will be used in the proof presented in Sec. \ref{sec4}.

To briefly summarise the different coordinates that one can have here, we have the canonical coordinates $\mathbf{z}_{c}$ on the coadjoint orbit in which the symplectic form may be written
\begin{align}
    \eta = J_{ij}dz_{c}^i \wedge dz_{c}^j,
\end{align}
the non-canonical coordinates $\mathbf{z}_n$ which are still local coordinates on the coadjoint orbit (so have the correct dimensionality), in which the symplectic form is instead
\begin{align}
    \eta = \eta_{ij}(\mathbf{z}_n) dz_n^i\wedge dz_n^j
\end{align}
and the Lie-Poisson coordinates $\mathbf{u}$ which describe the dynamics on the coadjoint orbit, but due to the larger dimensionality, a symplectic form associated with these coordinates cannot be defined in the same way.

Since the coherent state manifold $\mathbb{CP}^{K-1}$ is a symplectic leaf of $\mathfrak{su}(K)$, as we will show in Sec.~\ref{subsecChooseLeaf}, we wish to convert from the Lie-Poisson coordinates $\mathbf{u}$ to the local coordinates of $\mathbb{CP}^{K-1}$, $\theta^i$ and $\phi^i$, using the moment map. This results in the following expressions for $u_i$ in the cases of the symmetric, antisymmetric and diagonal GGM matrices:
\begin{subequations}\label{uFormulae}
\begin{align}
    u_{\alpha_{nm}}(\mathbf{\Omega}) &= 2\text{Re}(\braket{n}{\Omega} \overline{\braket{m}{\Omega}}), \\
    u_{\beta_{nm}}(\mathbf{\Omega}) &= 2\text{Im}(\braket{n}{\Omega} \overline{\braket{m}{\Omega}}), \\
    u_{\gamma_n}(\mathbf{\Omega}) &= \sqrt{\frac{2}{n(n-1)}}\left((1-n)(\braket{n}{\Omega}\overline{\braket{n}{\Omega}})+\sum_{k=1}^{n-1}\braket{k}{\Omega}\overline{\braket{k}{\Omega}}\right),
\end{align}
\end{subequations}
where $\bar{a}$ denotes the complex conjugate of $a\in\mathbb{C}$. The variables $u_i$ define an embedding of $\mathbb{CP}^{K-1}$ into $\mathbb{R}^{K^2-1}$ necessitated by the dimensionality of the Lie algebra. Although this increases computational cost with growing $K$, the vector $\mathbf{u}$ is the most natural for investigating the Poisson nature of the dynamics due to its connection with the KKS symplectic form through the Poisson bivector.

Using Eqn.~\eqref{CoherentState}, we can more explicitly calculate these dynamical variables by calculating
\begin{align}\label{CalculatingU}
    \braket{n}{\Omega} \overline{\braket{m}{\Omega}} &= \cos\frac{\theta^n}{2}\cos\frac{\theta^m}{2}\prod_{k=1}^{m-1}\sin^2\frac{\theta^k}{2} \prod_{l=m}^{n-1}\sin\frac{\theta^l}{2}e^{i\phi^l}.
\end{align}
which can be substituted back into Eqns.~\eqref{uFormulae} to find the angle dependence explicitly. It should be noted that while these angles are local coordinates, they are not the canonical local coordinates. Canonical local coordinates exist for $\mathbb{CP}^{K-1}$ in the form of the action-angle coordinates \cite{oh_action-angle_1994} but these use a different convention for $\phi$ than the spin-mapping literature, so we derive canonical coordinates for the spin mapping convention in Sec.~\ref{subsecCanonCoords}.

At this point, we remark that a symplectic method on $\mathbb{CP}^{K-1}$ must preserve Casimirs, but the inverse is not strictly true. A Casimir is defined as a function $C \in C^\infty(\mathfrak{g}^*)$ such that
\begin{align}
    \{F,C\}(u) = if_i{}_j{}^ku_k \frac{\partial F}{\partial u_i}\frac{\partial C}{\partial u_j} = 0 ~~\forall F \in C^{\infty}(\mathfrak{g}^*),u\in\mathfrak{g}^*.
\end{align}
Since the Casimirs are defined independently of the variable $u$ they do not capture local, $u$ dependent, effects that may be necessary to completely eliminate degeneracies within the Poisson bivector on $\mathbb{CP}^{K-1}$, so do not lead to a symplectic form. For this reason it is in general not sufficient to simply check the conservation of every Casimir in order to show symplecticity \cite{cook_spin-mint_2026}.

This can equivalently be seen by a dimension count. The $K^2-1$ dimensional $\mathfrak{su}(K)$ Lie algebra has $K-1$ independent Casimirs, thus the (real) dimension of the space which preserves all Casimirs is $K^2-K$, which is precisely the real dimension of the full flag. So a Casimir preserving method is symplectic if the orbit that the method propagates on is the full flag manifold. In contrast, the real dimension of the partial flag relative to this study, $\mathbb{CP}^{K-1}$ is $2K-2$ \cite{monk_geometry_1959} so we can clearly see that in general the two spaces cannot be equal. In fact the only case in which the two can be equated is precisely when $K^2-K = 2K-2$, so $K=2$. Since many of the spin systems appearing in literature use specifically $SU(2)$ \cite{lubich_symplectic_2010,bogfjellmo_collective_2018,ephrati_two_2025,mclachlan_symplectic_2014}, it is worthwhile to note this distinction that appears in the more general case. 

This distinction also matters in the spin-mapping case, where the notion of Casimirs extends trivially to the manifold $\mathbb{R}^{2F}\times \mathbb{CP}^{K-1}$. This is because the Casimirs of $\mathbb{R}^{2F}$ are precisely the constant functions, so Casimirs of the product space $\mathbb{R}^{2F}\times\mathbb{CP}^{K-1}$ are functions of the form $C_\times (R,P,u) = C(u)$ for $C$ a Casimir of $\mathfrak{su}(K)$, therefore the Casimirs are effectively unchanged by including the nuclear phase space. However, the symplectic form is changed to include the nuclear coordinates, introducing a new $(2F+2K-2)\times(2F+2K-2)$ symplectic structure matrix with couplings between canonical position coordinates on $\mathbb{CP}^{K-1}$ and canonical momentum coordinates on $\mathbb{R}^{2F}$ and vice versa. This non-trivial extension of the symplectic form shows that a method which is Casimir-preserving or even symplectic on $\mathbb{CP}^{K-1}$ may not be symplectic when extended to $\mathbb{CP}^{K-1}\times \mathbb{R}^{2F}$, even in the case $K=2$ \cite{kelly_mapping_2012,richardson_analysis_2017, cook_which_2023, cook_spin-mint_2026}. Recently, this was shown explicitly via the derivation of a Spin Split-Liouvillian (Spin-SL) algorithm which preserves the norm of the spin-vector but is not symplectic. \cite{cook_spin-mint_2026}

\subsection{The Spin-MInt Algorithm}

For propagation of the entire spin-mapping system, we must consider carefully the product space $\mathbb{R}^{2F}\times \mathbb{CP}^{K-1}$. We first apply the SW transform to the spin-mapping Hamiltonian from Eqn.~\eqref{Hamiltonian} which we then separate into two sub-Hamiltonians,
\begin{subequations}
\begin{align}
    H_1 (\mathbf{P}) &= \frac{1}{2} \mathbf{P}^\mathrm{T}\boldsymbol{\mu}^{-1}\mathbf{P}, \\
    H_2 (\mathbf{R},\mathbf{u}) &= \mathcal{H}_0(\mathbf{R}) + 2r_s\mathcal{H}^i(\mathbf{R})u_i.
\end{align}
\end{subequations}
Using these, the Spin-MInt algorithm is given by the symmetrically composed flow map \cite{cook_spin-mint_2026},
\begin{align}\label{SMFlowMap}
    \Phi_{\Delta t}^{\text{Spin-MInt}} = \phi_{H_1,\Delta t/2} \circ \phi_{H_2,\Delta t} \circ \phi_{H_1,\Delta t/2},
\end{align}
for sub-Hamiltonian flow maps $\phi_{H_i,t'}$ representing the numerical integration of Hamiltonian $H_i$ for a timestep $t'$.

The non-trivial coupling between the complex projective space and nuclear degrees of freedom present within the term $\mathcal{H}^i(\mathbf{R}) u_i$ means that both variables must be propagated simultaneously and exactly, as exact integration of sub-Hamiltonians is sufficient to guarantee a symplectic method \cite{leimkuhler_simulating_2005}. The proposed method for this propagation is the Spin-MInt algorithm given in \cite{cook_spin-mint_2026}, in which it was shown that the method is symplectic using conjugate canonical variables for two states and for an arbitrary number of electronic states via a comparison to a previously known symplectic method with different variables \cite{church_nonadiabatic_2018}. However, we look to present a direct explicit proof for a general number of electronic states which provides insight to fundamental geometry of the spin-mapping representation.

From the spin-mapping Hamiltonian, the key propagation equations are \cite{cook_spin-mint_2026}:
\begin{subequations}\label{eqn:PropEqns}
\begin{align}
    \dot{R}^k &= \frac{P^k}{\mu_{kk}}\\
    \dot{P}^k &= - \partial_k\mathcal{H}_0 - 2r_s \partial_k\mathcal{H}^i u_i\\
    \dot{u}_i &= f_i{}_j{}^l\mathcal{H}^ju_l := -iW_i{}^lu_l
\end{align}
\end{subequations}
where $\mu$ is the diagonal matrix of nuclear masses, $W_i{}^l := if_{ij}{}^l\mathcal{H}^j$, the indices are in the ranges $k \in \{1,\dots,F\}$ and $ i,j,l \in \{ 1,\dots,(K^2-1) \}$ and $\partial_k$ represents the derivative with respect to the $k$th nuclear position variable $R^k$. Since we are using Hamilton's equations for the nuclear variables here, one should include the canonical Poisson bivector $\delta^{kl}$, however, for visual simplicity, we omit this, hence the raised and lowered indices appearing inconsistent. We will continue to use this convention throughout the work. The Spin-MInt algorithm proposes the following updates of each of the variables from time $t$ to time $t+\Delta t$:
\begin{subequations}\label{eqn:PropEqns2}
\begin{align}
    R^k(t+\Delta t /2) &= R^k(t) + \frac{P_k(t)}{2\mu_{kk}}\Delta t, \\
    \label{PProp}P^k(t+\Delta t) &= P^k(t) -  \partial_k \mathcal{H}_0 \Delta t - 2r_s\int_t^{t+\Delta t} {\left[ e^{-i\mathbf{W}t'} \right]_{i}}^j \partial_k\mathcal{H}^i u_j(t) dt', \\
    u_i(t+\Delta t) &= {\left[ e^{-i\mathbf{W}\Delta t} \right]_{i}}^ju_j(t),\\
    R^k(t+\Delta t) &= R^k(t+\Delta t/2) + \frac{P_k(t+\Delta t)}{2\mu_{kk}}\Delta t.
\end{align}
\end{subequations}
The Hamiltonian components $\mathcal{H}_0$ and $\mathcal{H}_i$ as well as $\mathbf{W}$ seen in the $P^k$ and $u_i$ are all evaluated at position $\mathbf{R}(t+\Delta t/2)$ as this is what the position will be during the flow map $\phi_{H_2,\Delta t}$, having already undergone the flow map $\phi_{H_1,\Delta t/2}$. The integral appearing in Eqn.~\eqref{PProp} is integrated analytically via an eigenvector decomposition \cite{cook_spin-mint_2026}.

We note that showing this propagation is the exact propagation given by each sub-Hamiltonian is sufficient to show that the method is symplectic. However, again this approach fails to probe the true geometry of the system being considered and does not yield a monodromy matrix.

\section{Methodology}\label{sec3}

With this background theory covered, we are ready to properly join the mathematical formalism present in the study of the Lie-Poisson system through the dynamics on the dual space with the known structure of spin-mapping through previous work \cite{runeson_generalized_2020,runeson_spin-mapping_2019,bossion_non-adiabatic_2022,bossion_non-adiabatic_2023}. This is done by selecting the specific symplectic leaf $\mathbb{CP}^{K-1}$ on which we get the relevant spin-mapping dynamics. Since $\mathbb{CP}^{K-1}$ lends itself nicely to having canonical coordinates, we will find and use these as they will simplify the proof slightly. Here, we also construct the monodromy matrix which will be the main object of interest while working on the proof that the Spin-MInt algorithm is symplectic.

\subsection{Choosing A Leaf}\label{subsecChooseLeaf}

The general structure of all the symplectic coadjoint orbits of $\mathfrak{su}(K)$ is described by that of a flag manifold on the vector space $\mathbb{C}^{K}$. The flag manifold associated with $\mathfrak{su}(K)$ consists of a set of nested sub-manifolds of various even real dimensions \cite{monk_geometry_1959,arvanitoyeorgos_geometry_2006}. The full (largest dimensional) flag manifold has a dimension of $K(K-1)$, while the smallest non-trivial partial flag manifold $\mathbb{CP}^{K-1}$ has dimension $2K-2$. These submanifolds are precisely the coadjoint orbits that may be endowed with the KKS symplectic form, and thus the orbits on which we are able to consider dynamics \cite{Kirillov_Orbit_2004}. 

Recalling the isomorphism between $\mathfrak{su}(K)$ and $\mathfrak{su}(K)^*$ provided by the trace inner product Eqn.~\eqref{TrInnerProd}, we can therefore identify any $u \in \mathfrak{su}(K)^*$ with a matrix $\boldsymbol{\rho}\in\mathfrak{su}(K)$. This allows calculation of the individual $u_i$ variables via the moment map
\begin{align}\label{DualIsom}
    u_k = \text{Tr}[\boldsymbol{\rho}\boldsymbol{\lambda}_k].
\end{align}
If we enforce the condition that $\boldsymbol{\rho} = \ket{\Omega}\bra{\Omega}$ (often called a pure state \cite{sakurai_modern_2020}), then we retrieve the earlier moment map given in Eqn.~\eqref{PureMoMap} and pre-existing spin-mapping literature \cite{bossion_non-adiabatic_2021,bossion_non-adiabatic_2022,runeson_generalized_2020,runeson_spin-mapping_2019}, but so far, there is no reason to neglect the other allowed states (often called mixed states). Investigation of these mixed state is equivalent to investigating various larger partial flags. Therefore we look to motivate our choice of orbit resulting in the space $\mathbb{CP}^{K-1}$.

Fully understanding the general choice of coadjoint orbit requires the use of representation theoretical ideas \cite{humphreys_introduction_1972,perelomov_generalized_1986,atiyah_convexity_1982,delbourgo_maximum_1977,delbourgo_minimal_1977}, but since we use only the fundamental representation of $SU(K)$, we can instead proceed by considering the matrix nature of the coadjoint orbit $\mathcal{O}_u\subset \mathfrak{su}(K)^*$. This allows the classification of distinct coadjoint orbits by the eigenspectrum (the set of eigenvalues and associated multiplicities) of the density matrix \cite{marsden_introduction_1999}. To see how this can be done, we consult the definition of the coadjoint orbit given in Eqn.~\eqref{OrbDef}. For a point $v \in \mathcal{O}_u$ with associated matrix $\boldsymbol{\rho} \in \mathfrak{su}(K)$ via Eqn.~\eqref{DualIsom}, a point $v'$ with associated matrix $\boldsymbol{\rho}'$ is also in the coadjoint orbit $\mathcal{O}_u$ if 
\begin{align}\label{SameOrbit}
    \boldsymbol{\rho}' = g\boldsymbol{\rho}g^{-1}
\end{align}
for some $g \in SU(K)$, so the orbits of $SU(K)$ are precisely the conjugacy classes of the Hermitian matrices.

Two Hermitian matrices are conjugate by a unitary matrix if and only if they have the same eigenspectrum. Therefore the coadjoint orbits may be entirely classified as sets of matrices with equivalent eigenspectra. This reduces the problem of choosing a symplectic leaf down to picking the eigenspectrum of $\boldsymbol{\rho}$. Since $\boldsymbol{\rho}$ is a matrix playing the role of a state which can be propagated, it is a density matrix. Furthermore, since the $K$-level electronic system is closed, one wishes to use a density matrix defined by a pure state, so one of rank 1 \cite{sakurai_modern_2020}. The simplest rank 1 $K\times K$ matrix is $\boldsymbol{\rho} = \mathbf{E}_{11} = \ket{1}\bra{1}$, however, $\mathbf{E}_{11} \notin \mathfrak{su}(K)$ so we should in theory work with its projection onto the traceless space
\begin{align}
    \boldsymbol{\rho}_{0} = \begin{bmatrix}
        1-\frac{1}{K} & 0 & \dots & 0\\
        0 & -\frac{1}{K} & & \vdots\\
        \vdots & &\ddots &\\
        0 & \dots & & -\frac{1}{K}
    \end{bmatrix}.
\end{align}
In practice, however, since we will only look at its behaviour under the map $u_i = \text{Tr}[\boldsymbol{\rho} \boldsymbol{\lambda}_i]$, it is equivalent to consider the rank 1 operator $\boldsymbol{\rho}$ and its traceless projection $\boldsymbol{\rho}_0$ (which is rank $K$).

To find the coadjoint orbit that this state corresponds to, we have the moment map on the set of rank 1 operators: 
\begin{align}\label{Rank1Projection}
    u(\mathbf{X}) = \text{Tr}[\boldsymbol{\rho} \mathbf{X}] = \bra{1}\mathbf{X}\ket{1} = X_{11},
\end{align}
which in turn gives the associated stabiliser
\begin{align}
    \text{Stab}_{\mathfrak{su}(K)}(u) = \{ \mathbf{X} \in \mathfrak{su}(K) : X_{1i}Y_{i1} - Y_{1i}X_{i1} = 0 ~~ \forall \mathbf{Y}\in\mathfrak{su}(K) \}.
\end{align}
Since every element of $\mathfrak{su}(K)$ is skew-Hermitian, this can be rewritten as
\begin{align}
    \text{Stab}_{\mathfrak{su}(K)}(u) = \{ \mathbf{X} \in \mathfrak{su}(K) :  \text{Im}(X_{1i}Y_{i1}) = 0 ~~\forall \mathbf{Y} \in \mathfrak{su}(K) \}.
\end{align}
The only restriction on $\mathbf{Y}$ being that it must be skew Hermitian effectively allows all upper triangular elements other than the diagonals to be unconstrained, implying that for each $i\neq 1, X_{1i}$ must be zero. For $i=1, Y_{11} = i\alpha$ in general, so we can have $X_{11} = i\beta$ for any $\beta \in \mathbb{R}$. This results in $\mathbf{X}$ having the form
\begin{align}
    \mathbf{X} = \begin{bmatrix}
        i\beta & \mathbf{0}^\textrm{T} \\
        \mathbf{0} & \mathbf{A}
    \end{bmatrix}
\end{align}
for $\mathbf{A}$ a $(K-1)\times(K-1)$ skew-Hermitian matrix with $\text{Tr}(A) = -i\beta$ and $\mathbf{0}\in\mathbb{R}^{K-1}$. This is often written as $\mathbf{X}$ being an element of the Lie algebra $\mathfrak{s}(\mathfrak{u}(1) \times \mathfrak{u}(K-1))$. It is then known that the corresponding coadjoint orbit, $SU(K)/S(U(1)\times U(K-1))$ is isomorphic to $\mathbb{CP}^{K-1}$ \cite{leifer_cpn-1_2006}. The dimension of the coadjoint orbit may be calculated by subtracting the dimension of $\mathfrak{s}(\mathfrak{u}(1)\times\mathfrak{u}(K-1))$ from $K^2-1$. This gives $K^2-1 - (K-1)^2 = 2K-2$, resulting in the lowest dimensional non-trivial partial flag manifold.

\subsection{Dynamics in Lie-Poisson and Canonical Coordinates}\label{subsecCanonCoords}

To analyse the dynamics, one considers the time evolution of observables. We recall that this is done through the Poisson bracket acting on $\mathfrak{su}(K)^*$,
\begin{align}
    \dot{\mathbf{B}}(\mathbf{u}) = \{ i\mathbf{H},\mathbf{B} \}(\mathbf{u}) = f_i{}_j{}^k \mathcal{H}^{j}\mathcal{B}^iu_k.
\end{align}
This defines the Poisson bivector $(\pi_u)_{ij} = f_i{}_j{}^ku_k$ with which we can define a symplectic form. However, we can also define the symplectic form in local coordinates using the more standard definition of the Poisson bracket in classical mechanics
\begin{align}
    \dot{\mathbf{B}}(\mathbf{u}) = \{\mathbf{B},\mathbf{H}\}_{C}(\mathbf{u}) = \eta^{nm} \left( \frac{\partial \mathbf{u}(\mathbf{H})}{\partial \theta^m}\frac{\partial \mathbf{u}(\mathbf{B})}{\partial \phi^n}- \frac{\partial \mathbf{u}(\mathbf{H})}{\partial \phi^n}\frac{\partial \mathbf{u}(\mathbf{B})}{\partial \theta^m} \right)
\end{align}

Where $\{\cdot,\cdot\}_C$ represents a `classical' Poisson bracket defined on the coadjoint orbit rather than the Poisson bracket on the entire Lie-Poisson space. Due to the definition of the moment map, $\mathbf{u}(\mathbf{H})=\mathcal{H}^iu_i$. Since $\mathbf{B}$ and $\mathbf{H}$ are arbitrary, we can compare these two component-wise, giving the relationship
\begin{align}\label{nonCanonPropEqn}
    f_i{}_j{}^ku_k = \eta^{nm}\left( \frac{\partial u_i}{\partial \theta^m} \frac{\partial u_j}{\partial \phi^n}- \frac{\partial u_i}{\partial \phi^n} \frac{\partial u_j}{\partial \theta^m} \right)
\end{align}
where $\eta = \eta_{nm}d\theta^m \wedge d\phi^n$ is the symplectic form for $\mathbb{CP}^{K-1}$, often called the Fubini-Study form, adapted to the coordinates in the Spin-Mapping convention from the coordinates and Fubini-Study form given by Bengtsson \cite{bengtsson_geometry_2017}. We write the Poisson bivector (inverse of the symplectic form matrix) as $\eta^{nm}$ such that $\eta^{nm}\eta_{ml} = {\delta^n}_l$.

Using Eqns.~\eqref{nonCanonPropEqn} and \eqref{uFormulae}, we can find the matrix $\eta^{nm}$ and thus invert it, giving the following for elements of $\eta_{ij}$:
\begin{align}
    \eta_{ij} = \begin{dcases}
        0  &j<i,\\
        \cot\frac{\theta^i}{2}\prod_{l=1}^j\sin^2\frac{\theta^l}{2} &j\geq i.
    \end{dcases}
\end{align}
By Darboux's theorem for differential forms \cite{mcduff_introduction_2017}, there exists a coordinate system in which the symplectic form takes the canonical form. The upper triangular structure of $\eta_{ij}$ facilitates the explicit construction of such coordinates in this case. One finds that the variables $\phi^i$ are already canonical. To construct canonical coordinates to replace $\theta^i$, we first note that 

\begin{align}
    \cot \frac{\theta^i}{2} = \frac{\partial}{\partial \theta^i} \left( \log \left( \sin^2\frac{\theta^i}{2} \right) \right),
\end{align}
and so for the canonical coordinates $\Theta$, one can write the 1-form as
\begin{subequations}
\begin{align}
    d\Theta^j &= \sum_{i\leq j} \eta_{ij} d\theta^j\\
    &= \sum_{i\leq j} \prod_{l=1}^j \sin^2\frac{\theta^l}{2} \frac{\partial}{\partial \theta^i} \left( \log \left( \sin^2\frac{\theta^i}{2} \right) \right) d\theta^j\\\
    &= \prod_{l=1}^j \sin^2\frac{\theta^l}{2} d \left( \log \left( \prod_{i=1}^j \sin^2\frac{\theta^i}{2} \right)\right) \\
    &= d\left( \prod_{i=1}^j \sin^2\frac{\theta^i}{2} \right).
\end{align}
\end{subequations}
This shows that $\Theta^i = \prod_{l=1}^i \sin^2\frac{\theta^l}{2}$ is the canonical coordinate replacing $\theta^i$. A similar expression with an additional constant term and change in sign has been proposed in Ref.~\cite{bossion_non-adiabatic_2022}, and these vary only slightly from the usual action angle variables on $\mathbb{CP}^{K-1}$ \cite{oh_action-angle_1994}. Since these coordinates are canonical, they provide the standard forms of Hamilton's equations:

\begin{subequations}
\begin{align}
    \dot{\Theta}^i &= \frac{\partial \mathbf{H}}{\partial \phi^i},\\
    \dot{\phi}^i &= -\frac{\partial \mathbf{H}}{\partial \Theta^i},
\end{align}
\end{subequations}
which gives $\Theta$ the role as a position variable here, and similarly $\phi$ is now momentum. This is the opposite of how the variables were posed in the two state version of the Spin-MInt algorithm, however, this gives the standard form of the involution matrix when considering time reversal symmetry of the variables which is advantageous \cite{cook_spin-mint_2026}. Note that a $\delta^{ij}$ representing the Poisson bivector matrix in canonical coordinates has been omitted for clarity, but should be used to raise the lowered indices on the right hand sides. Using these new variables,the Poisson bracket is
\begin{align}\label{PoissonURelation}
    f_i{}^j{}_k u_j = \frac{\partial u_i}{\partial \phi^j}\frac{\partial u_k}{\partial\Theta^j} - \frac{\partial u_k}{\partial \Theta^j}\frac{\partial u_i}{\partial \phi^j}
\end{align}
where a sum over $j$ is implied on the right side, as again we omit the identity term $\delta^{jl}$. This form of the Poisson bracket is most useful for our calculations of elements of the monodromy matrix, but the antisymmetry of the structure constants may be exploited to make the RHS negative, making it more obvious that $\boldsymbol{\Theta}$ and $\boldsymbol{\phi}$ are playing the roles of position and momenta respectively relative to the normal Poisson bracket.

\subsection{Monodromy Matrix Definition}

Having established $\Theta^i$ and $\phi^i$ as canonical coordinates, we define the monodromy matrix

\begin{align}
    \mathbf{M}_{\textrm{SM}} = \begin{bmatrix}
    \mathbf{M}_{RR} & \mathbf{M}_{R\Theta} & \mathbf{M}_{RP} & \mathbf{M}_{R\phi}\\
    \mathbf{M}_{\Theta R} & \mathbf{M}_{\Theta \Theta} & \mathbf{M}_{\Theta P} & \mathbf{M}_{\Theta \phi}\\
    \mathbf{M}_{PR} & \mathbf{M}_{P\Theta} & \mathbf{M}_{PP} & \mathbf{M}_{P\phi}\\
    \mathbf{M}_{\phi R} & \mathbf{M}_{\phi\Theta} & \mathbf{M}_{\phi P} & \mathbf{M}_{\phi\phi}
    \end{bmatrix},
\end{align}
where $\mathbf{M}_{AB} = \partial \mathbf{A}(t+\Delta t) / \partial \mathbf{B}(t)$ is an $m \times n $ matrix where $\mathbf{A}\in \mathbb{R}^n$ and $\mathbf{B}\in\mathbb{R}^m$. If we can show that this matrix satisfies $\mathbf{M}_\mathrm{SM}\mathbf{J} \mathbf{M}_\mathrm{SM}^\textrm{T} = \mathbf{J}$, then we have shown the symplecticity of the algorithm  \cite{hairer_geometric_2006}, where $\mathbf{J}$ is the canonical structure matrix. $\mathbf{M}_\mathrm{SM}$ and $\mathbf{J}$ are both $2(F+K-1)\times 2(F+K-1)$ matrix for $K$ electronic states and $F$ nuclear degrees of freedom.

Since we separate the Hamiltonian into two parts, we have two monodromy matrices to calculate. The first of these describing the nuclear propagation is given by \cite{cook_which_2023, cook_spin-mint_2026, church_nonadiabatic_2018}

\begin{align}
    \mathbf{M}_{H_1} = \begin{bmatrix}
        \mathbf{I}_F & \mathbf{0} & \frac{1}{2}\boldsymbol{\mu}^{-1} & \mathbf{0}\\
        \mathbf{0} & \mathbf{I}_{K-1} & \mathbf{0} & \mathbf{0} \\
        \mathbf{0} & \mathbf{0} & \mathbf{I}_F & \mathbf{0} \\
        \mathbf{0} & \mathbf{0} & \mathbf{0} & \mathbf{I}_{K-1}
    \end{bmatrix},
\end{align}
where $\mathbf{I}_n$ is the $n\times n$ identity matrix, and one can infer the dimensions of the zero matrices by considering the identity matrices with which they share a column and row. Showing that $\mathbf{M}_{H_1}$ is symplectic is trivial via direct calculation using Eqn.~\eqref{sympCriterion}. Therefore, to show that the total flow map given in Eqn.~\eqref{SMFlowMap} is symplectic, all that remains to show that is the $\mathbf{M}_{H_2}$ is also symplectic.

The monodromy matrix $\mathbf{M}_{H_2}$ is more complicated as this is where the Lie-Poisson variables appear. To tackle this we need to relate the Lie-Poisson variables $\mathbf{u}$ to the canonical variables $\boldsymbol{\Theta},\boldsymbol{\phi}$. First, we require expressions for the derivatives of $\boldsymbol{\phi}(t+\Delta t)$ and $\boldsymbol{\Theta}(t+\Delta t)$ rather than just $\mathbf{u}(t+\Delta t)$ with respect to some arbitrary variable $f(t)$. For ease of notation we will take $\delta t = t + \Delta t$. Starting with $\boldsymbol{\phi}$, by considering two elements $u_{\alpha_{ji}}$ and $u_{\beta_{ji}}$ for some $i,j$, we find that
\begin{subequations}
\begin{align}
    \frac{\partial (\phi^i + \dots + \phi^{j-1}) (\delta t)}{\partial f(t)} &= \cos^2(\phi^i + \dots + \phi^{j-1})(\delta t) \frac{\partial \tan (\phi^i + \dots + \phi^{j-1})(\delta t)}{\partial f(t)}\\
    &= \frac{u_{\alpha_{ji}}^2(\delta t)}{u_{\alpha_{ji}}^2(\delta t) + u_{\beta_{ji}}^2(\delta t)} \frac{\partial \left(\frac{u_{\beta_{ji}}(\delta t)}{u_{\alpha_{ji}}(\delta t)}\right)}{\partial f(t)}\\
    &= \frac{u_{\alpha_{ji}}(\delta t) \frac{\partial u_{\beta_{ji}}(\delta t)}{\partial f(t)} - u_{\beta_{ji}}(\delta t) \frac{\partial u_{\alpha_{ji}}(\delta t)}{\partial f(t)}}{u_{\alpha_{ji}}^2(\delta t) + u_{\beta_{ji}}^2(\delta t)}
\end{align}
\end{subequations}
where we have used that $\partial x/\partial \tan(x) = \cos^2(x)$. Finding the derivative of a single variable $\phi^i$ is therefore equivalent to considering the case $j = i+1$ giving
\begin{align}\label{DerivPhi}
    \frac{\partial \phi^i(\delta t)}{\partial f(t)} = \frac{u_{\alpha(i)}(\delta t) \frac{\partial u_{\beta(i)}}{\partial f(t)} - u_{\beta(i)} \frac{\partial u_{\alpha(i)}}{\partial f(t)}}{u_{\alpha(i)}^2(\delta t) + u_{\beta(i)}^2(\delta t)} 
\end{align}
where $\alpha(i) = \alpha_{i+1,i}$ and $\beta(i) = \beta_{i+1,i}$.

Turning our attention to $\Theta$,
\begin{subequations}
\begin{align}
    u_{\gamma_i} &= \sqrt{\frac{2}{i(i-1)}} \left( (1-i) \cos^2\frac{\theta^i}{2} \prod_{l=1}^{i-1} \sin^2\frac{\theta^l}{2} + \sum_{k=1}^{i-1} \cos^2\frac{\theta^k}{2} \prod_{l=1}^{k-1}\sin^2\frac{\theta^l}{2} \right)\\
    &= \sqrt{\frac{2}{i(i-1)}} \left( (i-1) \prod_{l=1}^i \sin^2\frac{\theta^l}{2} - i \prod_{l=1}^{i-1}\sin^2\frac{\theta^l}{2} + 1 \right)\\
    &\label{uInTermsofTheta}= \sqrt{\frac{2(i-1)}{i}}\Theta^i - \sqrt{\frac{2i}{i-1}}\Theta^{i-1} + \sqrt{\frac{2}{i(i-1)}}.
\end{align}
\end{subequations}
Since $\Theta^K = 0$, we can directly calculate $\Theta^{K-1}$, which in turns gives $\Theta^{K-2}$, which iteratively gives expressions for all $\Theta^i$ in terms of $u$. This process results in the inductive hypothesis
\begin{align}
    \Theta^{K-i} = \sum_{l=1}^i \frac{K-i}{(K-l+1)(K-l)} - \frac{K-i}{\sqrt{2(K-l+1)(K-l)}}u_{\gamma_{K-l+1}}
\end{align}
which may be verified by considering $u_{\gamma_{K-i}}$ given by Eqn.~\eqref{uInTermsofTheta}. Rearranging this, we have
\begin{align}
    \Theta^{K-i-1} =& \left( \sum_{l=1}^i \frac{K-i-1}{(K-l+1)(K-l)} - \frac{K-i-1}{\sqrt{2(K-l+1)(K-l)}}u_{\gamma_{K-l+1}} \right)
    \\ &\nonumber+ \frac{1}{K-i} - \sqrt{\frac{K-i-1}{2(K-i)}}u_{\gamma_{K-i}}
\end{align}
so $\Theta^{K-(i+1)}$ is given by the appropriate formula whenever $\Theta^{K-i}$ is, and thus this iterative formula for $\Theta$ is valid for all $i$ such that $u_{\gamma_{K-i+1}}$ is defined, i.e. for all $i \in \{ 0,\dots,K-1 \}$. Now taking the derivative,
\begin{align}
    \frac{\partial \Theta^{K-i} (\delta t)}{\partial f(t)} = \sum_{l=1}^i \frac{K-i}{\sqrt{2(K-l+1)(K-l)}} \frac{\partial u_{\gamma_{K-l+1}}(\delta t)}{\partial f(t)}
\end{align}
or alternatively for $i \in \{1,\dots,K\}$
\begin{align}\label{DerivTheta}
    \frac{\partial \Theta^i(\delta t)}{\partial f(t)} = \sum_{l=1}^{K-i} \frac{i}{\sqrt{2(K-l+1)(K-l)}} \frac{\partial u_{\gamma(l)}(\delta t)}{\partial f(t)}
\end{align}
where $\gamma(l) = \gamma_{K-l+1}$. With these quantities calculated, we are now able to write the full monodromy matrix for the second part of the Hamiltonian in terms of $\mathbf{u}$.

To account for the fact that in the SW transform, the $\mathbf{u}$ variable is scaled by $2r_s$, we also scale $\Theta$ by $2r_s$ since it is linear in $\mathbf{u}$. This is not necessary for $\phi$ since $\phi$ scales as $\mathbf{u}^2/\mathbf{u}^2$.

Combining the derivatives calculated in Eqns.~\eqref{DerivPhi} and \eqref{DerivTheta}, we can state the monodromy matrix for the flow associated with $H_2$. Since each expression is complicated, we give each sub-matrix a label which is then associated to an expression later.
\begin{align}
    M_{H_2} = \begin{bmatrix}
        \mathbf{I}_F & \mathbf{0} & \mathbf{0} & \mathbf{0}\\
        \mathbf{a} & \mathbf{b} & \mathbf{0} & \mathbf{c} \\
        \mathbf{d} & \mathbf{e} & \mathbf{I}_F & \mathbf{f} \\
        \mathbf{g} & \mathbf{h} & \mathbf{0} & \mathbf{j}
    \end{bmatrix}.
    \label{eqn:h2mat}
\end{align}

Since $u_i(\delta t) = {[e^{-i\mathbf{W}\Delta t}]_{i}}^ju_j(t)$, the exponentiated matrix appears very frequently in these expressions. Therefore we adopt the shorthand $\mathbf{Q} = e^{-i\mathbf{W}\Delta t}, \mathbf{\check{Q}} = e^{-i\mathbf{W}t'}$ for $t'$ a dummy variable, while defining the symbols above \cite{cook_spin-mint_2026}. We will also use the convention (while defining these sub-matrices) of taking $i,j \in \{1,\dots,K-1\}$ and $k,l \in \{1,\dots,F\}$ so that it is obvious what the dimensions of the following sub-matrices are. The sub-matrices are given by the following expressions
\begin{subequations}\label{MonodromyElements}
\begin{align}
    a^i{}_k &= \frac{\partial \Theta^i(\delta t)}{\partial R^k(t+\Delta t/2)} = 2r_s\sum_{l=1}^{K-i}\frac{i}{\sqrt{2(K-l+1)(K-l)}}\partial_k Q_{\gamma(l)}{}^ju_{j}(t)\\
    b^i{}_j &= \frac{\partial \Theta^i(\delta t)}{\partial \Theta^j(t)} = \sum_{l=1}^{K-i}\frac{i}{\sqrt{2(K-l+1)(K-l)}}Q_{\gamma(l)}{}^m \frac{\partial u_{m}(t)}{\partial \Theta^j(t)}\\
    c^i{}_j &= \frac{\partial \Theta^i(\delta t)}{\partial \phi^j(t)} = 2r_s\sum_{l=1}^{K-i}\frac{i}{\sqrt{2(K-l+1)(K-l)}}Q_{\gamma(l)}{}^m\frac{\partial u_{m}(t)}{\partial \phi^j(t)}\\
    \nonumber d^k{}_l &= \frac{\partial P^k(\delta t)}{\partial R^l(t+\Delta t/2)} = - \Delta t \left( \partial_k\partial_l V_0+ \frac{1}{2}\text{Tr}\left(\partial_k\partial_l V\right) \right) \\
    & \hspace{0.85in}- 2r_s\int_{t}^{\delta t} \check{Q}_{i}{}^j \partial_k\partial_l\mathcal{H}^iu_{j}(t) + \partial_l \check{Q}_i{}^{j}\partial_k\mathcal{H}^iu_j(t) dt' \\
    e^k{}_i &= \frac{\partial P^k(\delta t)}{\partial \Theta^i(t)} = \int_t^{\delta t} \check{Q}_{j}{}^{m} \partial_k\mathcal{H}^j  \frac{\partial u_m(t)}{\partial \Theta^i(t)}dt'\\
    f^k{}_i &= \frac{\partial P^k(\delta t)}{\partial \phi^i(t)} = -2r_s \int_t^{\delta t} \check{Q}_{j}{}^{m} \partial_k\mathcal{H}^j \frac{\partial u_m(t)}{\partial \phi^i(t)}dt'\\
    g^i{}_k &= \frac{\partial \phi^i(\delta t)}{\partial R^k(t + \Delta t/2)} = \frac{u_{\alpha(i)}(\delta t) \partial_k Q_{\beta(i)}{}^ju_j(t) - u_{\beta(i)}(\delta t)\partial_k Q_{\alpha(i)}{}^ju_j(t)}{u_{\alpha(i)}^2(\delta t) + u_{\beta(i)}^2(\delta t)}\\
    h^i{}_j &= \frac{\partial \phi^i(\delta t)}{\partial \Theta^j(t)}= \frac{u_{\alpha(i)}(\delta t) Q_{\beta(i)}{}^m  \frac{ \partial u_m(t)}{\partial \Theta^j(t)} - u_{\beta(i)}(\delta t)Q_{\alpha(i)}{}^m \frac{\partial u_m(t)}{\partial \Theta^j(t)}}{2r_s \left(u_{\alpha(i)}^2(\delta t) + u_{\beta(i)}^2(\delta t)\right)}\\
    j^i{}_j &= \frac{\partial \phi^i(\delta t)}{\partial \phi^j(t)} = \frac{u_{\alpha(i)}(\delta t) Q_{\beta(i)}{}^m  \frac{ \partial u_m(t)}{\partial \phi^j(t)} - u_{\beta(i)}(\delta t)Q_{\alpha(i)}{}^m \frac{\partial u_m(t)}{\partial \phi^j(t)}}{u_{\alpha(i)}^2(\delta t) + u_{\beta(i)}^2(\delta t)}.
\end{align}
\end{subequations}
where we recall $\boldsymbol{\Theta}$ has been scaled up by $2r_s$ and that partial position derivatives $\partial_i$ are with respect to $R^i(t+\Delta t/2)$ as position is not propagated at all during this step.

\section{Results}\label{sec4}

Here we provide an outline of the remaining steps required to explicitly prove that the Spin-MInt algorithm is symplectic for any number of electronic states and nuclear degrees of freedom. We recall that we wish to show that the total flow map $\Phi_{\Delta t}^{\text{Spin-MInt}}$ is symplectic. To do this we wish to show that the individual flow maps $\phi_{H_1,\Delta t/2}$ and $\phi_{H_2\Delta t}$ are each symplectic. Due to the sparsity of the monodromy matrix $\mathbf{M}_{H_1}$, showing this is simple for the first flow map, but as we can see from Eqn.~\eqref{eqn:h2mat}, this is not the case for the second flow map. We will provide an outline of the remaining proof including the main tools arising from the Lie algebra that can be used for this explicit verification, but leave the full explicit calculation to Appendix \ref{secA1}.

\subsection{Proof}

Proving symplecticity is equivalent to showing that $\mathbf{M}_{H_2} \mathbf{J}^{-1} \mathbf{M}_{H_2}^\textrm{T} = \mathbf{J}^{-1}$ as indicated by Eqn.~\eqref{sympCriterion} for $\mathbf{J}$ the typical structure matrix for a $K-1+F$ dimensional canonical system. Using the form of $\mathbf{M}_{H_2}$ given in Eqn.~\eqref{eqn:h2mat}, we can evaluate the product
\begin{align}\label{SympCondMat}
    \mathbf{M}_{H_2} \mathbf{J}^{-1} \mathbf{M}_{H_2}^\textrm{T} = \begin{bmatrix}
        \mathbf{0} & \mathbf{0} & -\mathbf{I}_F & \mathbf{0}\\
        \mathbf{0} & \mathbf{cb}^\textrm{T}-\mathbf{bc}^\textrm{T} & \mathbf{ce}^\textrm{T}-\mathbf{a} -\mathbf{bf}^\textrm{T} & \mathbf{ch}^\textrm{T}-\mathbf{bj}^\textrm{T}\\
        \mathbf{I}_F & \mathbf{fb}^\textrm{T}-\mathbf{ec}^\textrm{T}+\mathbf{a}^\textrm{T} & \mathbf{fe}^\textrm{T}-\mathbf{d}-\mathbf{ef}^\textrm{T}+\mathbf{d}^\textrm{T} & \mathbf{fh}^\textrm{T}-\mathbf{ej}^\textrm{T}+\mathbf{g}^\textrm{T}\\
        \mathbf{0} & \mathbf{jb}^\textrm{T}-\mathbf{hc}^\textrm{T} & \mathbf{je}^\textrm{T}-\mathbf{g}-\mathbf{hf}^\textrm{T} & \mathbf{jh}^\textrm{T}-\mathbf{hj}^\textrm{T}
    \end{bmatrix}
\end{align}

The full algebraic details are presented in Appendix \ref{secA1}, but we will present the main tools within the proof here in order to provide insight as to how the Lie Poisson structure may be utilised to show certain structural relations about propagation of Lie-Poisson variables.

The main tool for approaching evaluation of the various products of sub-matrices will be the relationship provided by the equivalence of the two Poisson brackets, Eqn.~\eqref{PoissonURelation}. However, we will also require an identity relating $\partial \mathbf{Q}/\partial R^k$ to $\mathbf{Q}$ and $\int_{t}^{\delta t} \mathbf{\check{Q}} dt'$. We can find this by looking at the integral representation of the derivative of an exponentiated matrix (Duhamel's formula) \cite{wilcox_exponential_1967, cook_spin-mint_2026}. Defining $\mathbf{W}_k = \partial \mathbf{W}/\partial R^{k}$, we consider

\begin{subequations}
\begin{align}
    \mathbf{F}(t') :=& e^{-i\mathbf{W}(\delta t-t')}e^{-i(\mathbf{W}+\epsilon \mathbf{W}_k)(t'-t)}\\
    \implies \mathbf{F}(t) =& e^{-i\mathbf{W}\Delta t}, ~\mathbf{F}(\delta t) = e^{-i(\mathbf{W}+\epsilon \mathbf{W}_k)\Delta t},
    \\\frac{\partial \mathbf{F}}{\partial t'} =& e^{-i\mathbf{W}(\delta t-t')}(-i \epsilon \mathbf{W}_k) e^{-i(\mathbf{W}+\epsilon \mathbf{W}_k)(t'-t)}
\end{align}
\end{subequations}
Now using the fundamental theorem of calculus, 

\begin{align}
   e^{-i(\mathbf{W}+\epsilon \mathbf{W}_k)\Delta t} - e^{-i\mathbf{W}\Delta t} &= -i\epsilon\int_t^{\delta t} e^{-i\mathbf{W}(\delta t-t')}\mathbf{W}_k e^{-i(\mathbf{W}+\epsilon \mathbf{W}_k)(t'-t)}dt'
\end{align}
and letting $\epsilon\rightarrow 0$,
\begin{subequations}
\begin{align}
    \partial_k \mathbf{Q} &= -i\int_t^{\delta t} e^{-i\mathbf{W}(\delta t-t')}\mathbf{W}_k e^{-i\mathbf{W}(t'-t)}dt'\\
    &= -i\int_{0}^{\Delta t} e^{-i\mathbf{W}(\Delta t-t')} \mathbf{W}_k e^{-i\mathbf{W}t'} dt'\\
    \label{dQ/dR}\implies \partial_k Q_i{}^o&= Q_i{}^j \int_0^{\Delta t} (\check{Q}^\textrm{T})_j{}^lf_l{}_m{}^n \partial_k\mathcal{H}^m \check{Q}_n{}^odt',
\end{align}
\end{subequations}
since $\mathbf{\check{Q}}^\textrm{T} = e^{i\mathbf{W}t'}$.

Since $\mathbf{W}$ is the adjoint action of $\mathbf{V}$ in the Lie algebra \cite{cook_spin-mint_2026,hall_lie_2015}, the corresponding Lie group adjoint action $e^{i\mathbf{W}t}$ is related to the group adjoint map $\text{Ad}_\mathbf{g}(\mathbf{X}) = \mathbf{gXg}^{-1}$ for $\mathbf{g} = e^{i\mathbf{V}} \in SU(K)$, where $\mathbf{X} \in \mathfrak{su}(K)$. Since this conjugation map is a linear map on the Lie algebra, it can be represented using a single matrix, $\mathbf{Q}$. This is the definition of the adjoint representation of the Lie group. Elements of $\mathbf{Q}$ may be calculated by considering how the group adjoint map acts on the GGM matrices:
\begin{align}
    \mathbf{g}\boldsymbol{\lambda}_i\mathbf{g}^{-1} = Q_i{}^j\boldsymbol{\lambda}_j.
\end{align}
It can be shown for all Lie groups that this action preserves the Lie bracket \cite{hall_lie_2015},
\begin{align}\label{AdjAction}
    \mathbf{g}[\mathbf{X},\mathbf{Y}]\mathbf{g}^{-1} = [\mathbf{gXg}^{-1},\mathbf{gYg}^{-1}].
\end{align}
Therefore taking $\mathbf{X},\mathbf{Y}$ to instead be elements of the GGM matrices, for the left hand side of Eqn.~\eqref{AdjAction}
\begin{align}
    \mathbf{g}[\boldsymbol{\lambda}_i,\boldsymbol{\lambda}_j]\mathbf{g}^{-1} = if_i{}_j{}^k \mathbf{g}\boldsymbol{\lambda}_k\mathbf{g}^{-1} = if_i{}_j{}^kQ_k{}^l\boldsymbol{\lambda}_l,
\end{align}
and for the right hand side of Eqn.~\eqref{AdjAction},
\begin{align}
    [\mathbf{g}\boldsymbol{\lambda}_i\mathbf{g}^{-1},\mathbf{g}\boldsymbol{\lambda}_j\mathbf{g}^{-1}] = [Q_i{}^k\boldsymbol{\lambda}_k,Q_j{}^l\boldsymbol{\lambda}_l]= Q_i{}^kQ_j{}^l[\boldsymbol{\lambda}_k,\boldsymbol{\lambda}_l] = if_k{}_l{}^mQ_i{}^kQ_j{}^l\boldsymbol{\lambda}_m
\end{align}
Equating the left and right hand sides and evaluating component-wise in $\boldsymbol{\lambda}_m$, one finds
\begin{align}\label{CofactorRelations}
    f_i{}_j{}^k{Q_{k}}^m = f_k{}_l{}^mQ_i{}^kQ_j{}^l.
\end{align}

Now returning to eqn. \eqref{dQ/dR}, $\mathbf{\check{Q}}^\textrm{T}$ is also an element of the adjoint representation, so we can write use Eqn.~\eqref{CofactorRelations} on $\mathbf{\check{Q}}^\textrm{T}$ to give
\begin{subequations}
\begin{align}
    \partial_kQ_i{}^o &= Q_i{}^j \partial_k \mathcal{H}^m\int_0^{\Delta t} {\check{Q}}_j{}^{l} f_l{}_m{}^n  \check{Q}_{n}{}^{o}dt'\\
    &\label{ExpDerivNoBar}= -Q_{i}{}^j \partial_k\mathcal{H}^m \int_0^{\Delta t} f_j{}^o{}_l\check{Q}_{m}{}^{l} dt'.
\end{align}
\end{subequations}
Writing $\mathbf{\bar{Q}} = \int_0^{\Delta t} \mathbf{\check{Q}}dt'$, we have that
\begin{align}\label{ExpDerivEqn}
    \partial_kQ_i{}^j = -\partial_k\mathcal{H}^nQ_i{}^l\bar{Q}_n{}^mf_l{}^j{}_m.
\end{align}

The final set of relations that are most important for the calculations of monodromy matrix are explicit calculations of the structure constants. In the case of considering the $\mathfrak{su}(K)$ Lie algebra, the calculations were performed by considering the symmetric, antisymmetric and diagonal matrices as separate cases \cite{bossion_general_2021}. These calculations result in the following 
\begin{align*}
    f_{\alpha_{nm}\alpha_{kn}\beta_{km}} = f_{\alpha_{nm}\alpha_{nk}\beta_{km}} = f_{\alpha_{nm}\alpha_{km}\beta_{kn}} = f_{\beta_{nm}\beta_{km}\beta_{kn}} = 1,
\end{align*}
\begin{align}\label{ExpStrucCons}
    f_{\alpha_{nm}\beta_{nm}\gamma_m} = -\sqrt{\frac{2(m-1)}{m}},\ f_{\alpha_{nm}\beta_{nm}\gamma_n} = \sqrt{\frac{2n}{(n-1)}},
\end{align}
\begin{align*}
    f_{\alpha_{nm}\beta_{nm}\gamma_k} = \sqrt{\frac{2}{k(k-1)}}~~ \text{where}~m<k<n,
\end{align*}
and their antisymmetric permutations as the only non-zero structure constants of the $\mathfrak{su}(K)$ Lie algebra with the GGM basis as defined in Eqns.~\eqref{GGMBasis}. As far as we are aware, there does not exist a general method for the analytical calculation of the structure constants for an arbitrary Lie group $\mathfrak{g}$.

These relations are adequate to show that the matrix in Eqn.~\eqref{SympCondMat} is equal to $\mathbf{J}^{-1}$. Again due to the full calculation being rather notationally lengthy, from here we will provide an outline of how we have approached the proof, and refer the reader to Appendix \ref{secA1} for the entire process. 

We first look at the condition $-\mathbf{bj}^\textrm{T}+\mathbf{ch}^\textrm{T}$, where showing that this matrix is equal to  $-\mathbf{I}$ also gives a proof of the general satisfaction of Liouville's theorem by the flow map. More precisely, by showing this, we have implicitly shown that the determinant of the monodromy matrix $\mathbf{M}_{H_2}$ is unity, and therefore by Liouville's theorem that the top form of the manifold $\wedge^{K-1}\eta$ is conserved. This is done by considering separately the diagonal and off-diagonal elements, and showing using Eqn.~\eqref{ExpStrucCons} that these are 1 and 0 respectively.

We then move on to showing that the matrices on the block diagonal are 0. Each of these elements requires quite a different treatment to receive the desired result. $\mathbf{cb}^\textrm{T}-\mathbf{bc}^\textrm{T}=\mathbf{0}$ follows immediately from use of Eqn.~\eqref{CofactorRelations} and Eqn.~\eqref{ExpStrucCons}. For $\mathbf{jh}^\textrm{T}-\mathbf{hb}^\textrm{T}$, one uses the orbit preservation that we have shown earlier to rewrite $u(\delta t)$ in terms of arbitrary angle variables $\Tilde{\theta},\Tilde{\phi}$, where we can then use angle addition formulae to show the complete cancellation of each element. Finally, we show that the matrix $\mathbf{d}^\textrm{T}-\mathbf{d}+\mathbf{fe}^\textrm{T}-\mathbf{ef}^\textrm{T} = \mathbf{0}$ due to the antisymmetry of the structure constants. 

The final two constraints $\mathbf{ce}^\textrm{T} -\mathbf{a}-\mathbf{bf}^\textrm{T} = \mathbf{0}$ and $\mathbf{je}^\textrm{T}-\mathbf{g}-\mathbf{hf}^\textrm{T} = \mathbf{0}$ follow quickly from use of Eqn.~\eqref{PoissonURelation} and Eqn.~\eqref{ExpDerivEqn}.

\section{Discussion}\label{secUses}

\subsection{Formal Setting of Spin-Mapping}\label{sec51}

As far as we are aware, this is the first time that the formal mathematical setting for the spin-mapping representation has been presented, particularly motivating the form of the SW transform through consideration of the dual of the Lie algebra, and thoroughly investigating the usual choice of symplectic leaf $\mathbb{CP}^{K-1}$, including the symplectic form and canonical coordinates associated with the usual spin-mapping conventions. However, this formulation also permits the usage of density operators corresponding to mixed states which as far as we are aware have not yet been studied in the context of spin-mapping. The Spin-MInt propagation scheme should also give the symplectic propagation of the classical variables corresponding to mixed states due to using the Lie-Poisson $u$ variables, although care should be taken in establishing canonical variables on the larger partial flags that these mixed states correspond to if one wishes to construct monodromy matrices.

To give a brief outline of this, if one considered a mixed state density matrix which once diagonalised had two non-zero, equal eigenvalues
\begin{align}
    \boldsymbol{\rho} = \begin{bmatrix}
        \frac{1}{2} & 0 & 0 & \dots & 0\\
        0 & \frac{1}{2} & 0 & & 0\\
        0 & 0 & 0 & & 0\\
        \vdots & & & \ddots &\\
        0 & 0 & 0 & & 0
    \end{bmatrix},
\end{align}
with $u(\mathbf{X}) = \text{Tr}[\boldsymbol{\rho}\mathbf{X}]$ then the associated stabiliser becomes
\begin{align}
    \text{Stab}_{\mathfrak{su}(K)}(u) = \mathfrak{s}(\mathfrak{u}(2) \times \mathfrak{u}(K-2)).
\end{align}
This results in a coadjoint orbit $SU(K)/S(U(2)\times U(K-2)) \cong \mathbb{G}\text{r}_2(\mathbb{C}^K)$ the Grassmannian describing the complex space $\mathbb{C}^K$ where instead of having an equivalence relation on complex planes of real dimension 2 as in $\mathbb{CP}^{K-1}$, there is an equivalence relation on complex volumes of real dimension 4. This results in $\mathbb{G}\text{r}_2(\mathbb{C}^K)$ having a real dimension of $2\cdot 2\cdot(K-2) = 4(K-2)$. 

This idea may be further abstracted to consider any density matrix, where if it has a general diagonal form
\begin{align}
    \boldsymbol{\rho} = \begin{bmatrix}
        a_1 & 0 & \dots & 0\\
        0 & a_2 & & 0 \\
        \vdots & & \ddots &\\
        0 & 0 & & a_K
    \end{bmatrix}
\end{align}
with $a_1,a_2,\dots,a_K$ all distinct, then the resulting coadjoint orbit is the full flag $SU(K)/S(U(1)^K)$ with dimension $K(K-1)$. Cases where $a_i = a_j$ will reduce this dimension. A thorough investigation of the dynamics associated with and uses of the larger partial flags associated to the mixed states is left as future work.

\subsection{Application to the Semiclassical Initial Value Representation}

The Semiclassical Initial Value Representation (SC-IVR) is a very powerful method allowing for the effective numerical simulation of nonadiabatic systems. SC-IVR approximates the quantum propagator using the semiclassical formulation of quantum mechanics offered by the Hamilton-Jacobi equations, where the action is interpreted as a phase factor of the system \cite{zeng_development_2025}. There are many variants of SC-IVR \cite{miller_semiclassical_2001,ananth_semiclassical_2007,miller_electronically_2009,sun_semiclassical_1998}, but one with a significant success in terms of both accuracy and computational cost is the coherent state representation developed by Herman and Kluk \cite{herman_semiclasical_1984, kay_hermankluk_2006}. Loosely this proposes that for the MMST Hamiltonian $H$ with typical nuclear coordinates $R,P \in \mathbb{R}^F$ and MMST electronic coordinates $q,p\in\mathbb{R}^K$, one can reformulate the quantum propagator as follows
\begin{align}
    \nonumber e^{-iHt/\hbar} = \frac{1}{(2\pi\hbar)^{K+F}} &\iiiint_{\mathbb{R}^{2(K+F)}} C_t(P_0,R_0,p_0,q_0) e^{iS_t(P_0,R_0,p_0,q_0)/\hbar}\\
    & \cdot \ket{p_t,q_t}\bra{p_0,q_0} \otimes \ket{P_t,R_t}\bra{P_0,R_0} dP_0dR_0dp_0dq_0
\end{align}
where $x_t$ is the variable $x$ evaluated at time $t$ for $x\in\{P,R,p,q\}$, and similarly $S_t$ represents the classical action of the system at time $t$ given by
\begin{align}
    S_t(P_0,R_0,p_0,q_0) = \int_0^t P_{t'}\dot{R}_{t'} + p_{t'}\dot{q}_{t'} - H(P_{t'},R_{t'},p_{t'},q_{t'}) dt',
\end{align}
as implied by the Hamilton-Jacobi equations. $C_t$ is the so-called Herman-Kluk pre-exponential factor
\begin{align}\label{SCIVRPref}
    C_t(P_0,R_0,p_0,q_0) = \sqrt{\frac{1}{2}\left\lvert M_{\xi\xi}(t) + \Gamma^{-1}M_{\pi\pi}(t)\Gamma - i\hbar M_{\xi\pi}(t) \Gamma + \frac{i}{\hbar}\Gamma^{-1} M_{\pi\xi}(t) \right\rvert}.
\end{align}
where the matrices $M_{xy}(t)$ are defined as
\begin{align}
    \begin{bmatrix}
        \mathbf{M}_{\pi\pi}(t) & \mathbf{M}_{\pi\xi}(t)\\
        \mathbf{M}_{\xi\pi}(t) & \mathbf{M}_{\xi\xi}(t)
    \end{bmatrix}
    = \begin{bmatrix}
        \mathbf{M}_{PP}(t) & \mathbf{M}_{Pp}(t) & \mathbf{M}_{PR}(t) & \mathbf{M}_{Pq}(t)\\
        \mathbf{M}_{pP}(t) &\mathbf{M}_{pp}(t)&\mathbf{M}_{pR}(t)&\mathbf{M}_{pq}(t)\\
        \mathbf{M}_{RP}(t)&\mathbf{M}_{Rp}(t)&\mathbf{M}_{RR}(t)&\mathbf{M}_{Rq}(t)\\
        \mathbf{M}_{qP}(t)&\mathbf{M}_{qp}(t)&\mathbf{M}_{qR}(t)&\mathbf{M}_{qq}(t)
    \end{bmatrix}.
\end{align}
where $\mathbf{M}_{ij}(t) = \frac{\partial i(t)}{\partial j (0)}$ for $i,j\in\{P,R,p,q\}$. $\Gamma \in \mathbf{M}_{(F+K)\times(F+K)}$ is a diagonal matrix composed first of $F$ entries associated to the width of the nuclear wavepacket defined by the coherent state on $\mathbb{R}^{2F}$, and then $N$ entries associated to the width of the electronic wavepacket defined similarly. The explicit calculation of the monodromy matrix in the MMST based predecessor of the Spin-MInt algorithm, the MInt, has been used and adapted to provide explicit formulae for the calculation of the prefactor in Eqn.~\eqref{SCIVRPref} \cite{church_nonadiabatic_2018,moscato_time_2025}.

This formulation naturally leads to the question of whether a similar approach could be taken to SC-IVR within the spin-mapping representation. The present work does not seek to derive such a Spin-SC-IVR method, but simply highlights that semiclassical spin propagators based on spin coherent states have been proposed for spin-$j$ states \cite{novaes_semiclassical_2005,stone_semiclassical_2000}. These works also use elements of the monodromy matrix in their expressions of the semiclassical propagators, although they do not then go on to use the initial value representation as their work is not designed with nonadiabatic dynamics in mind. These propagators are proposed using non-canonical (stereographic \cite{bengtsson_geometry_2017,oh_action-angle_1994,barnes_cpn_2002}) coordinates for arbitrary representations of the $SU(2)$ Lie group rather than in local canonical coordinates of the $SU(K)$ Lie group, but we suspect that generalisation from the 2-level case to the $K$-level case should follow a similar structure to that taken within the spin-mapping literature, and that local canonical coordinates may be advantageous when it comes to calculation of such propagators. An example more comparable to the current setting where stereographic coordinates were utilised can be found in the study of system $(\mathbb{S}^2)^n \times T^*\mathbb{R}^m$ \cite{bogfjellmo_collective_2018}. If such a case were to arise, then the explicit calculations of monodromy matrices here would prove valuable to implementations of a Spin-SC-IVR method as was found to be the case with the MInt algorithm to MMST based approaches to SC-IVR \cite{moscato_time_2025}.

\subsection{Amendment of Non-Symplectic Methods}

Another advantage of an explicit solution is that one is actually able to analytically calculate a matrix $\mathbf{MJM}^\textrm{T}$, often called the error matrix \cite{cook_which_2023,cook_spin-mint_2026,church_nonadiabatic_2018} which in the case of a symplectic method should be identical to $\mathbf{J}$. This is a powerful diagnostic tool, as it enables one to investigate non-zero elements of the matrix $\mathbf{MJM}^\textrm{T}-\mathbf{J}$ and trace back through the calculation to investigate possible sources of lack of symplecticity. Investigation via this method may also highlight the changes one could make in order to amend an arbitrary method into a symplectic one. 

One could use the Frobenius norm as a metric for how close to symplectic an algorithm is \cite{cook_which_2023, cook_spin-mint_2026}, although caution should be exercised here since only in the case of an exactly symplectic method is this metric coordinate free. For example, if one defined a method which was a copy of an existing non-symplectic method on canonical coordinates $z$, but using the (also canonical but distinct) coordinates $z'$, then one would instead calculate the norm $\lVert M_{z'}JM_{z'}^\textrm{T} - J \rVert_F$ where $M_{z'} = \frac{\partial z'(\delta t)}{\partial z(\delta t)} M \frac{\partial z(t)}{\partial z'(t)}$. However, as is covered in literature on pseudo-symplectic methods \cite{aubry_pseudo-symplectic_1998,aubry_note_1998,stepanov_eight-stage_nodate}, the scaling of this error term as one alters the timestep $\Delta t$ is able to provide valuable information about the long time behaviour in deviation from the true value \cite{cook_spin-mint_2026, cook_which_2023}.


\section{Conclusion}\label{secConc}

In this paper we have discussed the formalism behind the symplecticity of numerical methods for simulating the Hamiltonian evolution of coherent states in Lie-Poisson systems. Using the ideas of pure and mixed quantum states, we have explored which symplectic sub-manifold of the Lie-Poisson manifold one should choose in order to design a symplectic method for classical-like propagation. The application of this to the spin-mapping system, very relevant in the space of nonadiabatic dynamics, was then explored. We then proved that the recently-proposed Spin-MInt algorithm \cite{cook_spin-mint_2026} is symplectic for any number of electronic states and nuclear degrees of freedom. 

Formalising the mathematical standing of the spin-mapping approach including the use of the coherent state to describe the SW transform provides valuable insight into the efficacy of spin-mapping. We believe that developing an understanding of the geometry of the manifold $\mathbb{CP}^{K-1}$ will help the design of current and future methods which look to employ the spin-mapping approach with larger numbers of states \cite{richardson_nonadiabatic_2025}. 

This investigation has also highlighted the canonical variables in which one can attain the cleanest version of the symplectic form, motivating their use over the complex stereographic coordinates used to study similar complex projective systems in the context of Hamiltonian dynamics. We hope that the calculations detailed here may provide a framework under which one could investigate the pseudo-symplecticity of a possibly cheaper method, enabling the quantisation of accumulation of errors. 

Furthermore, we would like to highlight the use of the Poisson structure within the proof of symplecticity. Using the structure constants and associated relations was of great benefit to the proof presented here, and they provide great value in the calculation of quantities such as the propagator in Lie-Poisson groups as well as finding alternate expressions for nuclear derivatives through the preservation of the Lie bracket under the conjugation action. For numerical implementations, calculations of the structure constants may be performed numerically. For analytical expressions, especially in the case of matrix Lie algebras whose basis elements are fairly sparse, explicit calculations of the commutator can be helpful \cite{bossion_general_2021} but in general there does not seem to be a good method for analytical calculation of structure constants.

Future work in this direction may include studying the relevance of the mixed states and their decoherence effects presented in Sec.~\ref{sec51}, and the development of SC-IVR methods compatible with the spin-mapping representation using the canonical form of the presented monodromy matrix.

\backmatter

\bmhead{Acknowledgements}
TJHH thanks Gregory Ezra for a helpful discussion. We acknowledge the use of the LLM `Claude' to assist in generating the tikz code for Fig.~\ref{fig:placeholder}.

\section*{Declarations}

\subsection*{Funding}
TJHH acknowledges a Royal Society University Research Fellowship URF\textbackslash R1\textbackslash 201502 and renewal URF\textbackslash R\textbackslash 251018. LEC acknowledges a University College London studentship. JRR acknowledges a summer studentship funded by TJHH's RS URF and funding from the Engineering and Physical Sciences Research Council [grant number EP/Z534882/1].
\subsection*{Competing Interests}
The authors declare that they have no competing interests.
\subsection*{Ethics approval and consent to participate}
Not applicable
\subsection*{Consent for publication}
Not applicable
\subsection*{Data availability}
Not applicable 
\subsection*{Materials availability}
Not applicable
\subsection*{Code availability}
Not applicable
\subsection*{Author contribution}
The algebraic derivation here was completed by JRR with guidance from LEC and supervision from TJHH. The initial draft manuscript was written by JRR and all authors contributed to editing and composing the final version.


\begin{appendices}

\section{Full Proof}\label{secA1}

Here, the full algebraic proof of symplecticity is presented by an explicit calculation of each element in the product matrix $\mathbf{M}_{H_2}\mathbf{J}^{-1}\mathbf{M}_{H_2}^\textrm{T}$ in Eq.~\eqref{SympCondMat}. Recalling the definitions of each monodromy matrix sub-matrix from Eqn.~\eqref{MonodromyElements}, we first start with trying to show that $\mathbf{ch}^\mathrm{T}-\mathbf{bj}^\mathrm{T} = -\mathbf{I}_K$.

\begin{align}
    &\nonumber[-\mathbf{bj}^\textrm{T}+\mathbf{ch}^\textrm{T}]^i{}_k = -b^i{}_jj_k{}^j +c^i{}_jh_k{}^j \\
    &= \sum_{l=1}^{K-i} \frac{i Q_{\gamma(l)}{}^p \left( Q_{\alpha(k)}{}^nu_n Q_{\beta(k)}{}^m  - Q_{\beta(k)}{}^nu_n Q_{\alpha(k)}{}^m \right)\left( \frac{\partial u_m}{\partial \phi^j} \frac{\partial u_p}{\partial \Theta^j} - \frac{\partial u_p}{\partial \phi^j} \frac{\partial u_m}{\partial \Theta^j}       \right)}{\sqrt{2(K-l+1)(K-l)} \left( (Q_{\alpha(k)}{}^ou_o)^2+(Q_{\beta(k)}{}^ou_o)^2 \right) },
\end{align}
which using eqn.~\eqref{PoissonURelation}
\begin{subequations}
\begin{align}
    &[-\mathbf{bj}^\textrm{T}+\mathbf{ch}^\textrm{T}]^i{}_k = \sum_{l=1}^{K-i} \frac{i Q_{\gamma(l)}{}^pf_m{}^j{}_pu_j \left( Q_{\alpha(k)}{}^nu_n Q_{\beta(k)}{}^m - Q_{\beta(k)}{}^nu_n Q_{\alpha(k)}{}^m \right)}{\sqrt{2(K-l+1)(K-l)} \left( (Q_{\alpha(k)}{}^ou_o)^2+(Q_{\beta(k)}{}^ou_o)^2 \right) }\\
    &=\sum_{l=1}^{K-i} \frac{i \left( Q_{\alpha(k)}{}^nu_n \left( -f_{\beta(k)\gamma(l)}{}^mQ_m{}^ju_j \right) - Q_{\beta(k)}{}^nu_n \left( -f_{\alpha(k)\gamma(l)}{}^mQ_m{}^ju_j \right)\right)}{\sqrt{2(K-l+1)(K-l)} \left( (Q_{\alpha(k)}{}^ou_o)^2+(Q_{\beta(k)}{}^ou_o)^2 \right) }\\
    &\label{OneBigFraction}=\sum_{l=1}^{K-i} \frac{i \left( Q_{\alpha(k)}{}^nu_n \left( -f^m{}_{\beta(k)\gamma(l)}Q_m{}^ju_j \right) - Q_{\beta(k)}{}^nu_n \left(f_{\alpha(k)}{}^m{}_{\gamma(l)}Q_m{}^ju_j \right)\right)}{\sqrt{2(K-l+1)(K-l)} \left( (Q_{\alpha(k)}{}^ou_o)^2+(Q_{\beta(k)}{}^ou_o)^2 \right) }.
\end{align}
\end{subequations}

Now we substitute the relevant values for the structure constants given in Eqn.~\eqref{ExpStrucCons}. Recalling that $\alpha(k),\beta(k)$ are referring to $\alpha_{k+1,k},\beta_{k+1,k}$, we can see that the numerator of eqn.~\eqref{OneBigFraction} will be 0 unless $K-l+1 = k$ or $k+1$ due to the indices for which the structure constants are non-zero given in Eqn.~\eqref{ExpStrucCons}. In the case $i=k$, since the sum is from $l=1$ to $l=K-i$, $K-l+1 = i$ is impossible, and $K-l+1 = i+1$ is obtained only when $l=K-i$. Therefore, the diagonal elements of $-\mathbf{bj}^\textrm{T}+\mathbf{ch}^\textrm{T}$ may be written as
\begin{align}
    [-\mathbf{bj}^\textrm{T}+\mathbf{ch}^\textrm{T}]^i{}_i &=\frac{-i \sqrt{\frac{2(i+1)}{i}} \left( (Q_{\alpha(i)}{}^nu_n)^2 + (Q_{\beta(i)n}{}^nu_n)^2     \right)}{\sqrt{2(i+1)i} \left( (Q_{\alpha(i)}{}^ou_o)^2+(Q_{\beta(i)}{}^ou_o)^2 \right) } = -1
\end{align}

In the case $i\neq k$, there needs to exist $l\in \{1,\dots K-i\}$ s.t. $K-k+1 = l$ or $K-k = l$. Therefore, if $i > k$, this is impossible and so $f=0 ~~\forall i,k$ meaning the total sum is 0. If $i<k$, then both $l=K-k$ and $l=K-k+1$ terms appear, such that
\begin{subequations}
\begin{align}
    \nonumber[-\mathbf{bj}^\textrm{T}+\mathbf{ch}^\textrm{T}]^i{}_k =& \frac{-i\left(Q_{\alpha(k)}{}^nu_n ( f^m{}_{\beta(k)\gamma_{k+1}} Q_m{}^ju_j) + Q_{\beta(k)}{}^nu_n (f_{\alpha(k)}{}^m{}_{\gamma_{k+1}} Q_m{}^ju_j) \right)}{\sqrt{2(k+1)k} \left( (Q_{\alpha(k)}{}^o u_o)^2 + (Q_{\beta(k)}{}^ou_o)^2 \right)}\\
    &\hspace{-0.3in}+ \frac{-i\left(Q_{\alpha(k)}{}^nu_n ( f^m{}_{\beta(k)\gamma_{k}} Q_m{}^ju_j) + Q_{\beta(k)}{}^nu_n (f_{\alpha(k)}{}^m{}_{\gamma_{k}} Q_m{}^ju_j) \right)}{\sqrt{2k(k-1)} \left( (Q_{\alpha(k)}{}^o u_o)^2 + (Q_{\beta(k)}{}^ou_o)^2 \right)}\\
    =& \frac{-i \sqrt{\frac{2(k+1)}{k}}}{\sqrt{2(k+1)k}} + \frac{i \sqrt{\frac{2(k-1)}{k}}}{\sqrt{2k(k-1)}} = \frac{-i}{k} + \frac{i}{k} = 0.
\end{align}
\end{subequations}
Therefore we have shown that $-\mathbf{bj}^\textrm{T} + \mathbf{ch}^\textrm{T} = -\mathbf{I}$, as desired for the symplecticity matrix and by taking the negative transpose, $-\mathbf{hc}^\textrm{T}+\mathbf{jb}^\textrm{T} = \mathbf{I}$. This also shows that the determinant of the monodromy matrix is unity. This further gives the satisfaction of Liouville's theorem such that the spin system's phase space volume is constant under propagation of the algorithm.

We now look at the elements on the diagonal of $\mathbf{M}_{H_2}\mathbf{J}^{-1}\mathbf{M}_{H_2}^\textrm{T}$. Unfortunately, there is no symmetry property which one can immediately to exploit to show that each of these elements is 0. We start with explicit verification for $-\mathbf{bc}^\textrm{T}+\mathbf{cb}^\textrm{T}$ as this is the easiest. Similar to before, we can identify the Poisson bracket term, yielding the expression
\begin{align}
    [\mathbf{cb}^\textrm{T} - \mathbf{bc}^\textrm{T}]^i{}_k = 2r_s f_m{}^j{}_ou_j \sum_{l=1}^{K-i}\sum_{n=1}^{K-k} \frac{i Q_{\gamma(l)}{}^mkQ_{\gamma(n)}{}^o}{\sqrt{4(K-l+1)(K-l)(K-n+1)(K-n)}} ,
\end{align}
where using Eqn.~\eqref{CofactorRelations}, we will only have terms of the form $f_{\gamma(l)\gamma(k)m}$ which using the structure constants, Eqns.~\eqref{ExpStrucCons}, is 0.

Turning our attention to $-\mathbf{hj}^\textrm{T}+\mathbf{jh}^\textrm{T}$, we have
\begin{subequations}
\begin{align}
    \nonumber [\mathbf{jh}^\textrm{T}-\mathbf{hj}^\textrm{T}]^i{}_k =& \frac{1}{2r_s} f_m{}^j{}_ou_j \Bigg( \frac{Q_{\alpha(i)}{}^lu_l Q_{\beta(i)}{}^m - Q_{\beta(i)}{}^lu_l Q_{\alpha(i)}{}^m}{(Q_{\alpha(i)}{}^pu_p)^2 + (Q_{\beta(i)}{}^p u_p)^2} \\ 
    &\hspace{0.5in}\cdot \frac{Q_{\alpha(k)}{}^nu_n Q_{\beta(k)}{}^o - Q_{\beta(k)}{}^nu_n Q_{\alpha(k)}{}^o}{(Q_{\alpha(k)}{}^qu_q)^2 + (Q_{\beta(k)}{}^q u_q)^2} \Bigg)\\
     &\hspace{-1in} = -u_m\frac{ u_{\alpha(i)}u_{\alpha(k)}f_{\beta\beta m} - u_{\alpha(i)}u_{\beta(k)}f_{\beta\alpha m} - u_{\beta(i)}u_{\alpha(k)}f_{\alpha\beta m} + u_{\beta(i)}u_{\beta(k)}f_{\alpha\alpha m}}{2r_s\left( (Q_{\alpha(i),p}u_p)^2 + (Q_{\beta(i),p} u_p)^2 \right)\left((Q_{\alpha(k),q}u_q)^2 + (Q_{\beta(k),q} u_q)^2\right)},
\end{align} 
\end{subequations}
where the indices on $f$ always relate $i$ to the first index and $k$ to the second, i.e. $f_{\alpha \beta m} = f_{\alpha(i)\beta(k)m}$, and each $u$ variable is evaluated at time $\delta t$, so are given by, for example, $u_{\alpha(i)} = Q_{\alpha(i)}{}^ju_j$. We find that all of these structure constants are only non-zero when $i+1 = k$ in the case when $k>i$, which is sufficient to consider since we know the matrix is antisymmetric. Explicitly calculating these, the numerator is
\begin{align}
    \nonumber(-u_{\alpha_{i+1,i}}u_{\alpha_{i+2,i+1}}u_{\beta_{i+2,i}} + u_{\alpha_{i+1,i}}u_{\beta_{i+2,i+1}}u_{\alpha_{i+2,i}}
    \\ + u_{\beta_{i+1,i}}u_{\alpha_{i+2,i+1}}u_{\alpha_{i+2,i}} + u_{\beta_{i+1,i}}u_{\beta_{i+2,i+1}}u_{\beta_{i+2,i}} ),
\end{align}
with each $u$ evaluated at time $\delta t$. Since the coadjoint orbit is preserved by unitary transforms from Eqn.~\eqref{SameOrbit}, and $e^{-i\mathbf{W}\Delta t}$ is a unitary transform, we know that $u(\delta t) \in \mathbb{CP}^{K-1}$ and therefore there exist some variables $\Tilde{\theta}^j(\delta t),\Tilde{\phi}^j(\delta t)$ such that $u(\delta t)$ can be calculated with these variables via Eqns.~\eqref{uFormulae}. Therefore, using Eqn.~\eqref{CalculatingU}, we can rewrite this numerator (again at time $\delta t$) as
\begin{align}
    \nonumber&\Gamma \Big( -\cos\Tilde{\phi}^i \cos\Tilde{\phi}^{i+1}\sin(\Tilde{\phi}^{i}+\Tilde{\phi}^{i+1}) + \cos\Tilde{\phi}^i \sin\Tilde{\phi}^{i+1}\cos(\Tilde{\phi}^{i}+\Tilde{\phi}^{i+1})\\
    & + \sin\Tilde{\phi}^i \cos\Tilde{\phi}^{i+1}\cos(\Tilde{\phi}^{i}+\Tilde{\phi}^{i+1}) + \sin\Tilde{\phi}^i \sin\Tilde{\phi}^{i+1}\sin(\Tilde{\phi}^{i}+\Tilde{\phi}^{i+1}) \Big),
\end{align}
where $\Gamma$ is a prefactor depending on $\Tilde{\theta}$. Expanding using the standard angle addition formulae results in cancellation, giving $[\mathbf{jh}^\textrm{T}-\mathbf{hj}^\textrm{T}]^i{}_k = 0 ~ \forall i<k$, which, by the antisymmetry of the matrix, yields the general result $\mathbf{jh}^\textrm{T}-\mathbf{hj}^\textrm{T}=0$.

\begin{figure}
    \centering
    \resizebox{\linewidth}{!}{\begin{tikzpicture}[
    scale=2.6,
    every node/.style={font=\small},
    axis/.style={-{Latex[length=2mm]}, thick},
    domainline/.style={thick},
    diag/.style={dashed, thick, gray}
]

\begin{scope}[xshift=0cm]
    \fill[blue!25] (0,0) rectangle (1,1);

    \draw[->] (-0.15,0) -- (1.2,0) node[right] {$s$};
    \draw[->] (0,-0.15) -- (0,1.2) node[above] {$t'$};

    \draw[domainline] (0,0) rectangle (1,1);

    \node[below left] at (0,0) {$0$};
    \node[below] at (1,0) {$\Delta t$};
    \node[left] at (0,1) {$\Delta t$};

    \node at (0.5,1.6) {\textbf{Square:} $0\leq s,t'\leq \Delta t$};
\end{scope}

\node at (1.6,0.5) {\Large $-$};

\begin{scope}[xshift=2.1cm]
    \fill[red!30] (0,0) -- (1,0) -- (1,1) -- cycle;

    \draw[->] (-0.15,0) -- (1.2,0) node[right] {$s$};
    \draw[->] (0,-0.15) -- (0,1.2) node[above] {$t'$};

    \draw[domainline, gray] (0,0) rectangle (1,1);

    \draw[diag] (0,0) -- (1,1);

    \draw[domainline] (0,0) -- (1,0) -- (1,1) -- cycle;

    \node[below left] at (0,0) {$0$};
    \node[below] at (1,0) {$\Delta t$};
    \node[left] at (0,1) {$\Delta t$};

    \node at (0.5,1.6) {\textbf{Triangle:} $0\leq s\leq t'\leq \Delta t$};
\end{scope}

\node at (3.8,0.5) {\Large $=$};

\begin{scope}[xshift=4.4cm]
    \fill[green!35!black!20!green] (0,0) -- (0,1) -- (1,1) -- cycle;

    \draw[->] (-0.15,0) -- (1.2,0) node[right] {$s$};
    \draw[->] (0,-0.15) -- (0,1.2) node[above] {$t'$};

    \draw[domainline, gray] (0,0) rectangle (1,1);

    \draw[diag] (0,0) -- (1,1);

    \draw[domainline] (0,0) -- (0,1) -- (1,1) -- cycle;

    \node[below left] at (0,0) {$0$};
    \node[below] at (1,0) {$\Delta t$};
    \node[left] at (0,1) {$\Delta t$};

    \node at (0.5,1.6) {\textbf{Complement ($\Delta$):} $0\leq t'\leq s\leq \Delta t$};
\end{scope}

\end{tikzpicture}}
    \caption{The domain of integration resulting from the subtraction of the triangular domain (red) from the square domain (blue), resulting in the complementary triangular domain (green), which is denoted $\Delta$.}
    \label{fig:placeholder}
\end{figure}
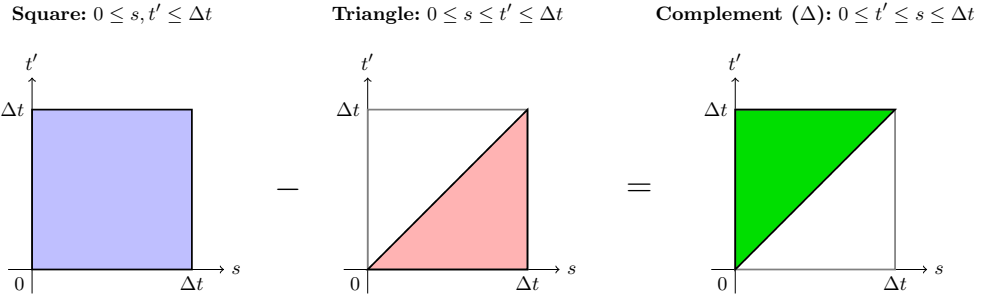

The final diagonal element to consider is $-\mathbf{d}-\mathbf{ef}^\textrm{T}+\mathbf{d}^\textrm{T}+\mathbf{fe}^\textrm{T}$. Proceeding as before,

\begin{align}
    \nonumber[-\mathbf{d}-\mathbf{ef}^\textrm{T}+\mathbf{d}^\textrm{T}+\mathbf{fe}^\textrm{T}]^k{}_l =& 2r_s \int_{0}^{\Delta t} \left( \partial_l \check{Q}_i{}^j \partial_k\mathcal{H}^i - \partial_k \check{Q}_i{}^j\partial_k\mathcal{H}^i \right)u_jdt' \\
    &+ 2r_sf_m{}^i{}_o u_i \int_{0}^{\Delta t}\check{Q}_{j}{}^m\partial_k\mathcal{H}^jdt'\int_{0}^{\Delta t}\check{Q}_{n}{}^o\partial_l\mathcal{H}^ndt',
\end{align}
using Eqn.~\eqref{ExpDerivNoBar} and choosing to explicitly write $\check{Q}$ to avoid confusion with integration variable labels, the right hand side is,
\begin{align}
    2r_sf_m{}^j{}_ou_j \Bigg( &\left( \partial_l\mathcal{H}^n\partial_k\mathcal{H}^i-\partial_l\mathcal{H}^i\partial_k\mathcal{H}^n\right) \int_0^{\Delta t} {\left[e^{-i\mathbf{W}t'}\right]_{i}}^o\int_{0}^{t'}{\left[e^{-i\mathbf{W}s}\right]_n}^mdsdt' \\
    &+ \partial_l\mathcal{H}^i\partial_k\mathcal{H}^n \int_{0}^{\Delta t}\int_0^{\Delta t}{\left[e^{-i\mathbf{W}t'}\right]_i}^o{\left[e^{-i\mathbf{W}s}\right]_n}^mdsdt' \Bigg).
\end{align}
Inspection of the integration domains shows that the first term corresponds to integration over the triangular domain $0\leq s\leq t'\leq \Delta t$, while the second spans the square domain $0\leq s,t' \leq \Delta t$. Subtracting the former from the latter leaves the complementary triangular domain $\Delta$ with $t',s$ such that $0\leq t' \leq s \leq \Delta t$, as seen in Fig. \ref{fig:placeholder}. From here, relabeling integration variables and dummy variables that are summed over such that all integrals are over the domain $\Delta$, the right hand side may be rewritten as
\begin{align}
    2r_sf_m{}^j{}_ou_j  \partial_l\mathcal{H}^n \partial_k\mathcal{H}^i  \int_\Delta {\left[e^{-i\mathbf{W}t'}\right]_i}^o{\left[ e^{-i\mathbf{W}s}\right]_n}^m + {\left[e^{-i\mathbf{W}s}\right]_n}^o{\left[e^{-i\mathbf{W}t'}\right]_i}^m ds dt'.
\end{align}
Finally, noting that the sum within the integral is symmetric in $o,m$ and that $f$ is antisymmetric, by relabeling $o\leftrightarrow m$ we can see that this expression is equal to its own negative, and therefore is equal to 0. 

This leaves only two more constraints to look at,
\begin{subequations}
\begin{align}
    C^i{}_k &= -a^i{}_k - b^i{}_j(f^j{}_k)^\textrm{T} +c^i{}_j(e^j{}_k)^\textrm{T} =0,\\
    D^i{}_k &\label{SympCondEqn}= -g^i{}_k-h^i{}_j(f^j{}_k)^\textrm{T} + j^i{}_j(e^j{}_k)^\textrm{T} = 0,
\end{align}
\end{subequations}
which we will investigate in a similar way. 
These two conditions are very similar, only distinguished by reversing the roles of $\Theta$ and $\phi$ in each, such that showing one is the same approach for the second. To verify Eqn.~\eqref{SympCondEqn}, it is equivalent to show that $g^i{}_k = -h^i{}_jf_k{}^j + j^i{}_je_k{}^j$, where by writing these terms out explicitly and noticing the common denominator, 
\begin{align}
    &\nonumber Q_{\alpha(i)}{}^lu_l \partial_k Q_{\beta(i)}{}^ju_j - Q_{\beta(i)}{}^lu_l \partial_k Q_{\alpha(i)}{}^j u_j =\\ 
    &\bar{Q}_{no}\partial_k \mathcal{H}^n\left(Q_{\alpha(i)}{}^pu_p Q_{\beta(i)}{}^m - Q_{\beta(i) }{}^p u_p Q_{\alpha(i)}{}^m\right)\left(\frac{\partial u_m}{\partial \Theta^j} \frac{\partial u_o}{\partial \phi^j} - \frac{\partial u_m}{\partial \phi^j} \frac{\partial u_o}{\partial \Theta^j}\right).
\end{align}

Now using Eqn.~\eqref{PoissonURelation} for the right hand side and Eqn.~\eqref{ExpDerivEqn} for the left hand side, we wish to show that
\begin{align}
    &\nonumber\bar{Q}_{n}{}^o\partial_k\mathcal{H}^n \left(-Q_{\alpha(i) }{}^mu_mQ_{\beta(i)}{}^lf_l{}^j{}_ou_j + Q_{\beta(i) }{}^mu_m  Q_{\alpha(i)}{}^l f_l{}^j{}_ou_j\right)=\\
    & \bar{Q}_n{}^o\partial_k\mathcal{H}^n \left( -Q_{\alpha(i) }{}^pu_p Q_{\beta(i) }{}^m f_m{}^j{}_ou_j + Q_{\beta(i)}{}^pu_p Q_{\alpha(i)}{}^mf_m{}^j{}_o u_j \right).
\end{align}
Quickly noting that all the differing indices are summed over in the same ways, it follows that $D^i{}_k =0$. By taking the negative transpose, we can also see that the condition on $-D^\textrm{T}$ will be zero, and following a similar procedure both $C$ related conditions can be shown to be 0. Therefore, the Spin-MInt algorithm is symplectic for all $K,F$.




\end{appendices}



\end{document}